\documentclass[superscriptaddress,twocolumn,showpacs,preprintnumbers,amsmath,
amssymb,prb]{revtex4}

\usepackage{color}
\usepackage{graphicx}
\usepackage{dcolumn}
\usepackage{bm}
\usepackage{subfigure}

\begin{document}


\title{Elastic transport through dangling-bond silicon wires on H passivated Si(100)}

\author{Mika\"el Kepenekian}
\email{mikael.kepenekian@cin2.es}
\affiliation{Centre d'Investigaci{\'o} en Nanoci{\`e}ncia i Nanotecnologia
       (ICN-CSIC), UAB Campus, E-08193 Bellaterra, Spain.}

\author{Frederico D. Novaes}
\affiliation{Centre d'Investigaci{\'o} en Nanoci{\`e}ncia i Nanotecnologia
       (ICN-CSIC), UAB Campus, E-08193 Bellaterra, Spain.}

\author{Roberto Robles}
\affiliation{Centre d'Investigaci{\'o} en Nanoci{\`e}ncia i Nanotecnologia
       (ICN-CSIC), UAB Campus, E-08193 Bellaterra, Spain.}

\author{Serge Monturet}
\affiliation{Nanoscience Group \& MANA satellite, CEMES/CNRS, 29 rue J. Marvig,
 BP 4347, 31055 Toulouse, C\'edex, France.}

\author{Hiroyo Kawai}
\affiliation{Institute of Materials Research and Engineering, 3 Research Link,
 Singapore 117602, Singapore.}

\author{Christian Joachim}
\affiliation{Nanoscience Group \& MANA satellite, CEMES/CNRS, 29 rue J. Marvig, 
 BP 4347, 31055 Toulouse, C\'edex, France.}

\author{Nicol\'as Lorente}
\affiliation{Centre d'Investigaci{\'o} en Nanoci{\`e}ncia i Nanotecnologia
       (ICN-CSIC), UAB Campus, E-08193 Bellaterra, Spain.}

\date{\today}

\begin{abstract}
	We evaluate the electron transmission through a dangling-bond wire on
	Si(100)-H (2$\times1$). Finite wires are modelled by decoupling semi-infinite
	Si electrodes from the dangling-bond wire with passivating H atoms. The
	calculations are performed using density functional theory in a non-periodic
	geometry along the conduction direction. We also use Wannier functions to
	analyze our results and to build an effective tight-binding Hamiltonian that
	gives us enhanced insight in the electron scattering processes. We evaluate
	the transmission to the different solutions that are possible for the
	dangling-bond wires: Jahn-Teller distorted ones, as well as antiferromagnetic
	and ferromagnetic ones. The discretization of the electronic structure of the
	wires due to their finite size leads to interesting transmission properties
        that are fingerprints of the wire nature.
        
\end{abstract}



\maketitle

\section{Introduction}

The development of information technology has been pursued at a tremendous
pace. Larger capacity memories and faster processors are obtained from the
miniaturization of electronic devices. Nevertheless, this technological explosion
that started in the second half of the 20$^{th}$ century will reach a limit when
facing the nanometer scale. Indeed, at this size, the bulk properties of
semiconductors are not accessible any more. Thus, the operating principles 
will vanish along their miniaturisation process.~\cite{muller1999a, meindl2001a}
{ In the mid-1970s, an alternative was proposed that single molecules
could perform the same basic functions of electronics devices that are traditionaly
silicon-based fabricated.~\cite{aviram1974a, joachim2000a} }
For this purpose, organic molecules are candidates with great potential given the
control on molecular design rending possible by chemical synthesis. In particular,
one can conceive and experiment on single molecules that switch from one state to 
another under the application of some external stimulus~\cite{kahn1988a} or 
even amplify a signal.~\cite{joachim2000a} Binary data can also be encoded in a
single molecule to build up complex circuit. Then, logic operations, either basic 
(AND, NOT, OR) or more elaborated can be performed at the molecular level, 
giving rise to molecular logic gates.~\cite{raymo2002a, liu2010a, puntoriero2011a,
renaud2011b}

\begin{figure}[!ht]
	\begin{center}
		\includegraphics*[scale=.58]{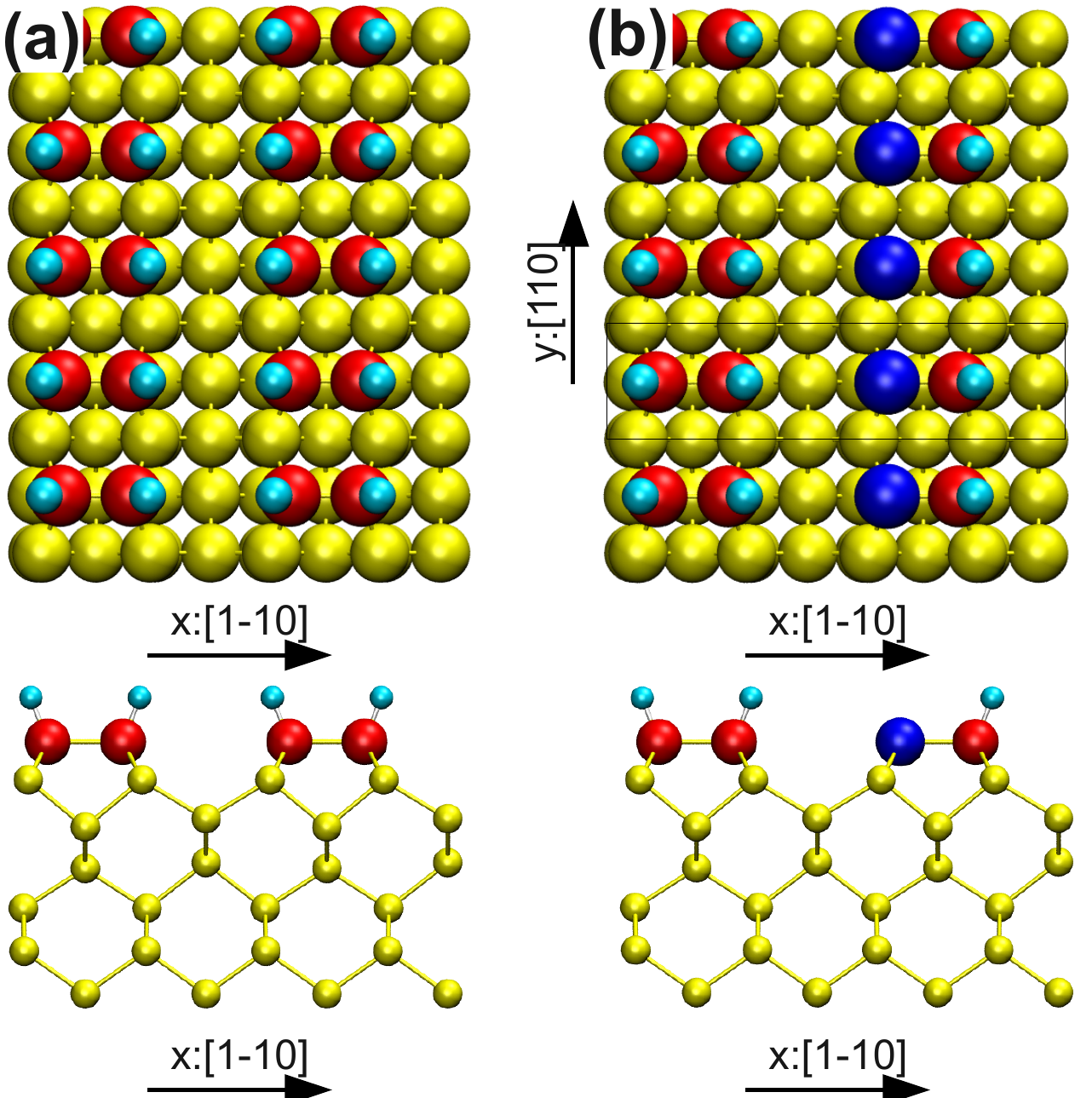}
	\end{center}
	\vspace{-0.3cm}
	\caption{Atomic structure of (a) the Si(100)-(2x1)-H surface, (b) the infinite
		ideal wire drawn along the $y$ direction. H atoms are depicted in
		cyan, while Si atoms are depicted in red (surface dimers), yellow
		(others) and blue when holding a dangling-bond.}
	\label{fig:db_wire}
\end{figure}
One interesting problem is to design the atomic scale circuit where a number of
these molecular logic gates will be (i) assessed and (ii) interconnected to obtain
more complex circuits. {Well designed atomic scale circuits can also
perform logic functions by themselves without the need of molecular logic
gates.~\cite{kawai2012a}}
It has been shown that such atomic scale structures can be constructed using the
scanning tunneling microscope (STM). It can for example selectively remove H
atoms on the surface of H-passivated semiconductors to construct dangling bonds
(DBs)lines.~\cite{shen1995a, hosaka1995a, hitosugi1999a, soukiassian2003a,
hallam2007a, haider2009a, pitters2011a}
Those DBs are introducing electronic states in the surface band gap of the passivated
semiconductor surface. An important example of such construction is the DB wire
produced by the selective removal of hydrogen atoms from an H-passivated Si(001)
surface along the Si dimer row (see Fig.~\ref{fig:db_wire}).
This DB wire has a single dangling bond per Si atom, intuitively offering a 1D metallic
band within the gap. The transport electronic properties of such a wire has been
previously inspected by Extended H{\"u}ckel calculations and explained using a
tight-binding model.~\cite{doumergue1999a, kawai2010a}
Unfortunately, the later configuration is not stable and a Peierls distortion
takes place, involving a metal-insulator transition. The unstable DB wire,
referred to as the ideal wire in the following, can also relax in two magnetic
forms resulting from the antiferromagnetic and ferromagnetic coupling of
adjacent DBs, respectively. The description of these different states has been
the subject of intense activity thanks to density functional theory (DFT) based
calculations.~\cite{watanabe1996a, watanabe1997a, cho2002a, bird2003,
cakmak2003a, lee2008a, lee2009a, lee2011a, nous2011a}
In particular, when dealing with finite-size wires, the magnetic solutions 
appear to be more stable than the distorted surface structure, 
referred to as the non-magnetic (NM) wire.~\cite{bird2003,lee2009a, nous2011a}
Although these relaxed periodic structures break the metallicity of the DB wire
by opening a gap, it does not dispose of the possibility to find a pseudo ballistic
or a tunnel transport regime through a finite length DB wire.

In the present work we inspect the transport properties of finite-size DB wires 
(ideal, NM, AFM and FM) using the non-equilibrium Green's Function (NEGF)
approach combined with ab initio DFT.
In order to complete the first model proposed by Kawai {\it et al.},~\cite{kawai2010a}
this study is preceded by a thorough analysis of the infinite ideal wire by means
of DFT-based calculations and Maximally Localized Wannier Functions
(MLWFs).~\cite{marzari1997a}

\section{Computational Details}

First-principles calculations are based on density functional theory (DFT) as
implemented in {\sc Siesta}.~\cite{soler2002a, artacho2008a} Calculations have been
carried out with the GGA functional in the PBE form,~\cite{perdew1996a}
Troullier-Martins pseudopotentials,~\cite{troullier1991a} and a basis set of finite-range
numerical pseudoatomic orbitals for the valence wave functions.~\cite{artacho1999a}
Structures have been relaxed using a double-$\zeta$ polarized basis
sets,~\cite{artacho1999a} while the conductance has been computed from
first-principles, using a single-$\zeta$ polarized basis set, by means of the {\sc TranSiesta}
method,~\cite{brandbyge2002a} within the non-equilibrium Green's function (NEGF)
formalism. In all cases, an energy cutoff of 200 Ry for real-space mesh size has
been used.

In order to reach a better understanding of the transport properties of the DB silicon
wires, subsequent transformation of the Kohn-Sham orbitals to Maximally Localized
Wannier Functions (MLWFs)~\cite{marzari1997a} has been applied.
This scheme allows one to obtain a localized orthogonal basis sets. As a consequence
MLWFs offer an extremely convenient way to translate the problem in terms of an
orthogonal tight-binding approach. Plus, an ab initio evaluation of the on-site
energies and hopping integrals becomes available.
The numerical calculations of MLWF have been run with the {\sc Wannier90}
code,~\cite{mostofi2008a} used as a post-processing tool of {\sc Siesta}. The interface
between both codes has been developed earlier by
R. Koryt{\'a}r {\it et al.}.~\cite{korytar2010a, korytar2011a}

\section{Ideal wire and H-junctions}
\label{sec:ideal}

The fully hydrogenated Si(100) surface presents a (2x1) reconstruction with
dimer rows formed along the $y$ ({\it i.e.} [110]) direction (see Fig.~\ref{fig:db_wire}
(a)). Starting from this fully passivated surface, one can construct a DB wire by removing
H atoms along the dimer rows direction (see Fig.~\ref{fig:db_wire} (b)). Before its relaxation,
this DB wire presents a metallic character. Although this ideal DB wire relaxes in either a
distorted wire (NM wire) or a magnetic phase (AFM and FM wires), the understanding of its
electronic transmission properties remains a capital point as other cases will be regarded
with respect to this reference.

\begin{figure}[!ht]
	\begin{center}
		\includegraphics*[scale=.55]{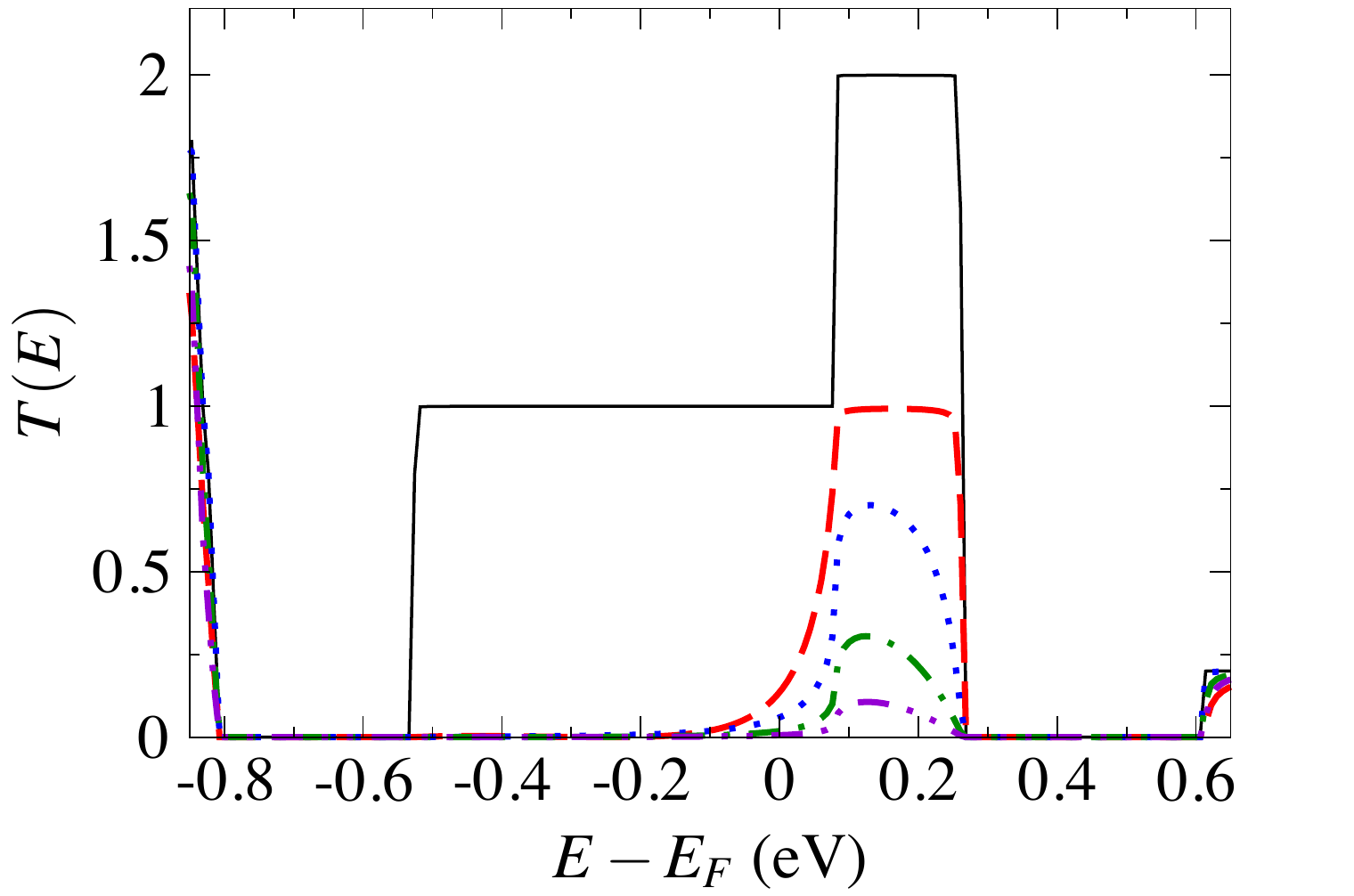}
	\end{center}
	\vspace{-0.5cm}
	\caption{Transmission as a function of energy through the ideal wire (black solid
		line) and 1 (dashed red liner), 2 (blue dotted line), 3 (green dotted and
		dashed line) and 4 (purple double dotted and dashed line) H-tunneling
		junctions.}
	\label{fig:transmission_ideal}
\end{figure}

Figure~\ref{fig:transmission_ideal} shows the transmission of an infinite ideal wire. As
expected from its band structure, the transmission exhibits clear steps featuring the
existence of two channels between 0.08 eV and 0.26 eV above the Fermi energy
($E_F$), with only one remaining channel for $E - E_F$ taken between $-0.53$ and $0.08$
eV. This result slightly differs  from the one previously obtained by Kawai
{\it et al.}~\cite{kawai2010a} from Extended H{\"u}ckel calculations, where the energy
range of the one-channel location is as large as the two-channel one ($\sim 0.4$ eV), the total
no zero transmission in the surface gap being the same in both calculations.

\begin{figure}[!ht]
	\begin{center}
		\includegraphics*[scale=.25]{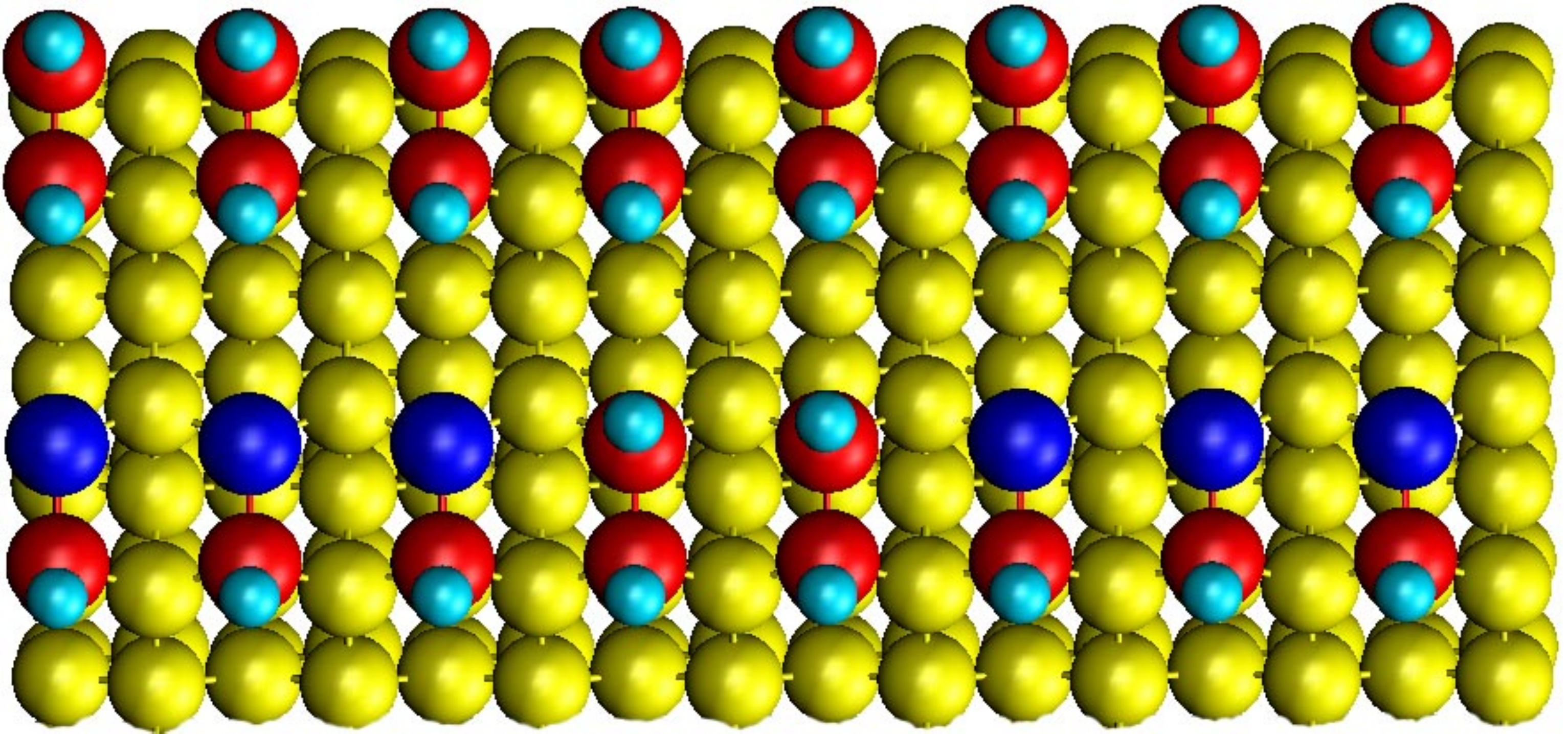}
	\end{center}
	\vspace{-0.3cm}
	\caption{Example of a 2H-junction. H atoms are depicted in cyan. Si atoms are
		depicted in red (surface dimers), yellow (others) and blue when holding
		a dangling-bond.}
	\label{fig:h_structure}
\end{figure}

The intuitive picture is that a transmission through such these non-relaxed DB wire arises
from the direct through space coupling between DBs. However, the substrate is  thought
to play an important role in the transport,~\cite{kawai2010a} and a convenient way to
\emph{get rid} of the direct coupling is to insert an H-passivated dimer in between two
semi-infinite ideal DB wires (see Fig.~\ref{fig:h_structure}). This peculiar structure will be
considered in section~\ref{sec:model} to get a better understanding of the transport and
scattering mechanisms in these systems.  More passivated dimers can be added in
between  the two ideal DB wires to track the tunnel decay through such an atomic scale
surface tunnel junction.

The transmission through different length H-junctions is presented
Figure~\ref{fig:transmission_ideal}. In the case of H-junctions longer than one
passivated dimer, a tunneling regime is observed in the middle of the gap with a
transmission $T(E)$ decreasing exponentially with the length of the H-junction.
For a chosen energy, the transmission $T(E)$ can be described as:
$$
T(E) = T_0 \, e^{-\gamma(E) d},
$$
where $d$ is the distance between semi-infinite ideal DB wires in angstrom and
$\gamma$ the inverse tunnel decay length (\AA$^{-1}$). At the Fermi energy, the
inverse decay length is found to be $\gamma = 0.29$~\AA$^{-1}$ ; a value
comparable to the one 0.22~\AA$^{-1}$ calculated with the Extended H{\"u}ckel
semi-empirical approximation.~\cite{kawai2010a}
\begin{figure}[!ht]
	\begin{center}
		\includegraphics*[scale=.5]{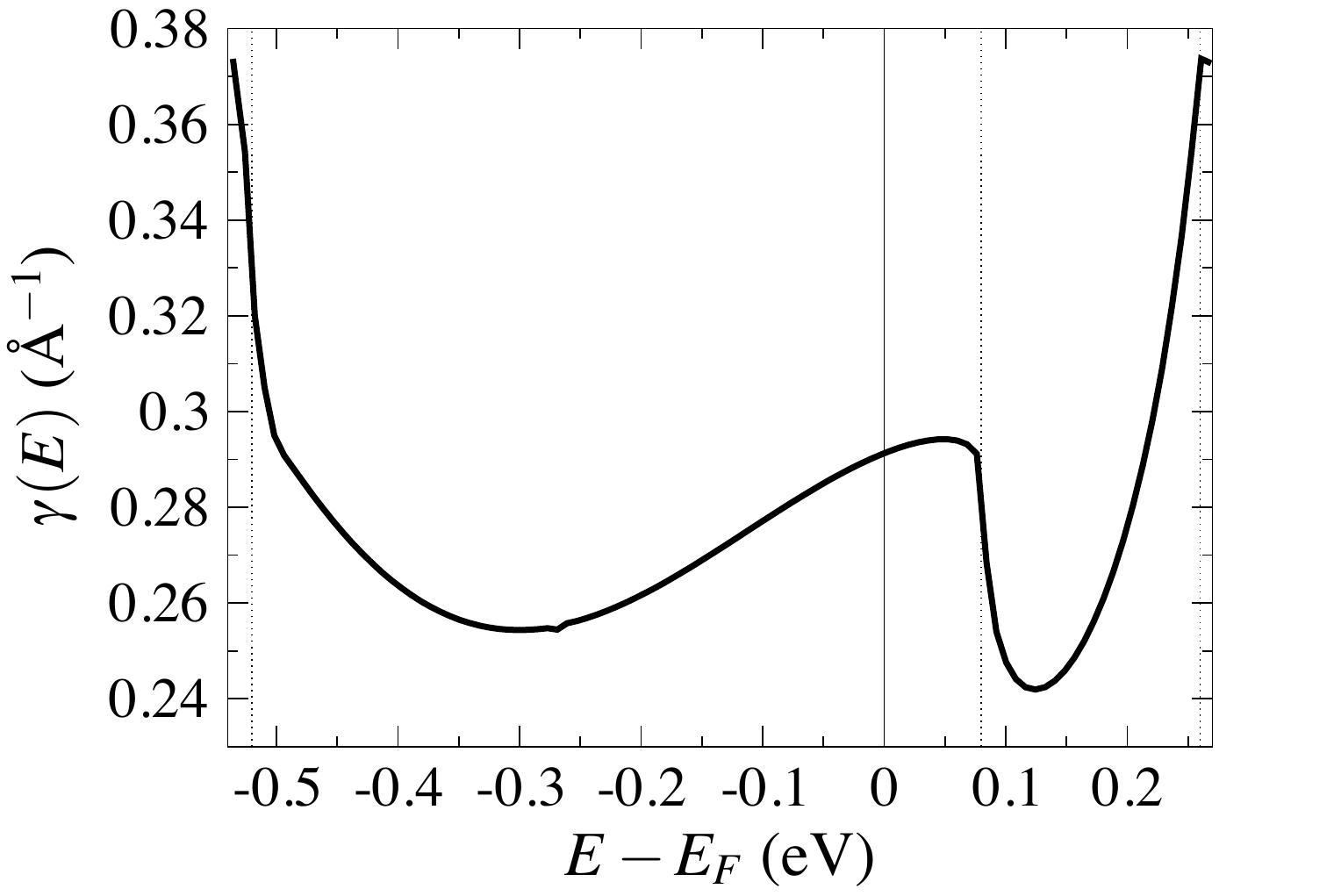}
	\end{center}
	\vspace{-0.5cm}
	\caption{Variation of the tunnel decay length $\gamma (E)$ (\AA$^{-1}$) as a
		function of electron incident energy. Dashed lines correspond to the
		energies of the steps in the transmission of the ideal DB wire.
		}
	\label{fig:gamma}
\end{figure}
$\gamma$ can be monitored with respect to the energy (see Fig.~\ref{fig:gamma}). It
lies between 0.24 and 0.37~\AA$^{-1}$, and exhibits two local minima (maxima for the decay
length Fig.~\ref{fig:gamma} curve) corresponding each to the center of the one channel
and two channels energy range of the ideal DB wire transmission spectrum (see
Fig.~\ref{fig:transmission_ideal}).

The situation is qualitatively different for a single H-passivated dimer inserted between
the semi-infinite wires. Indeed, this specific junction removes only one of the channels
and does not affect the second channel (active between 0.08 and 0.26 eV).

At this point, the ab initio DFT-treatment, although powerful, is too global to give a
clear picture of the transport mechanism through surface DB wires and H-junctions. 
In this regard, the parametrized tight-binding method constitutes an elegant way to model the
physics of transport. Thus, one would like to take the best of both worlds and
associate the accuracy of DFT-based calculations with the clarity of a simple tight-binding
basis set. This is made possible by the use of Wannier functions, which allows one to
(i) identify the main couplings, (ii) extract ab initio evaluations of on-site energies
and hoping integrals, and (iii) use them in a tight-binding calculation.

\section{Wannier functions' analysis}

\begin{figure}[!ht]
	\begin{center}
		\includegraphics*[height=3.1cm]{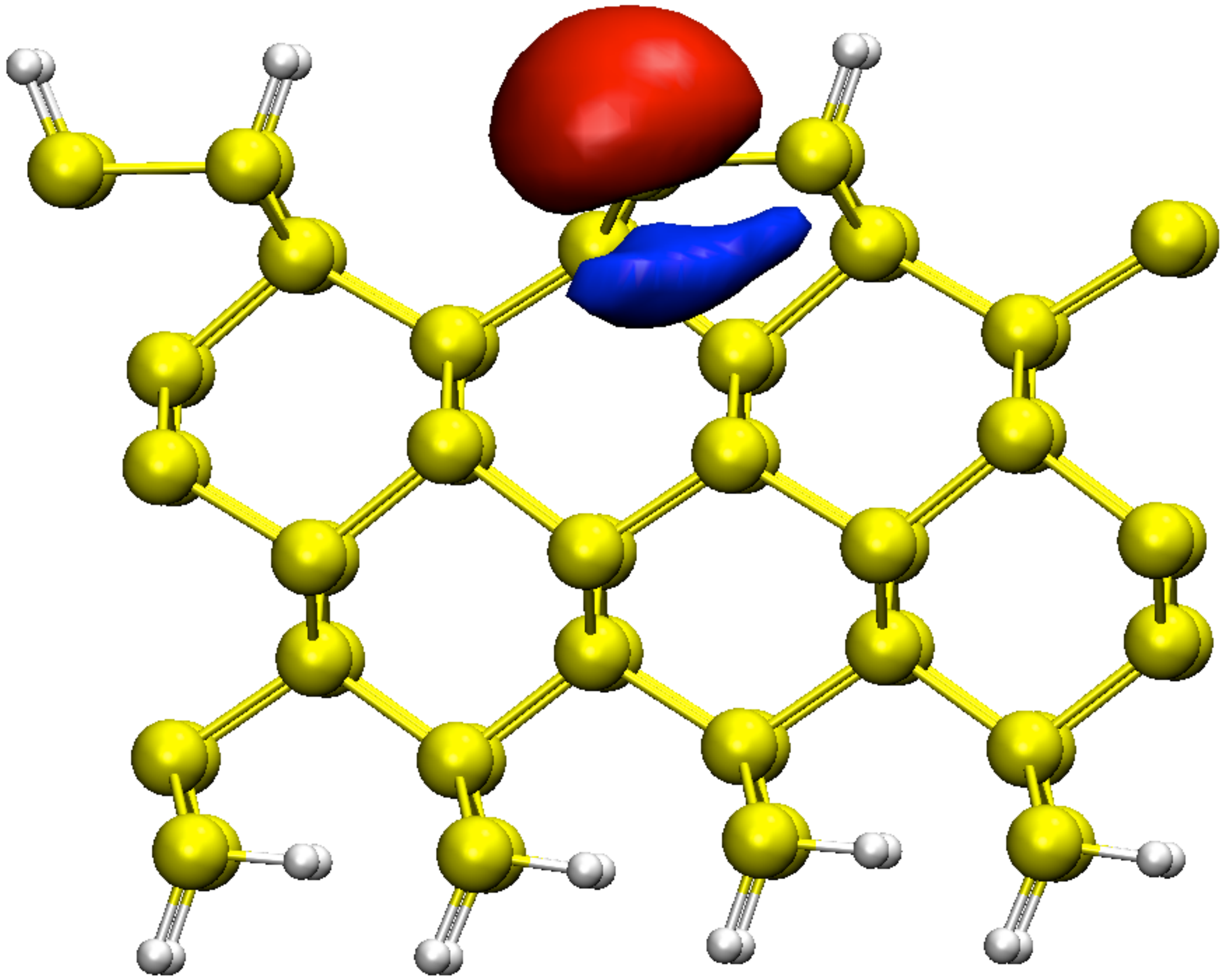}
		\hspace{.4cm}
		\includegraphics*[height=3.1cm]{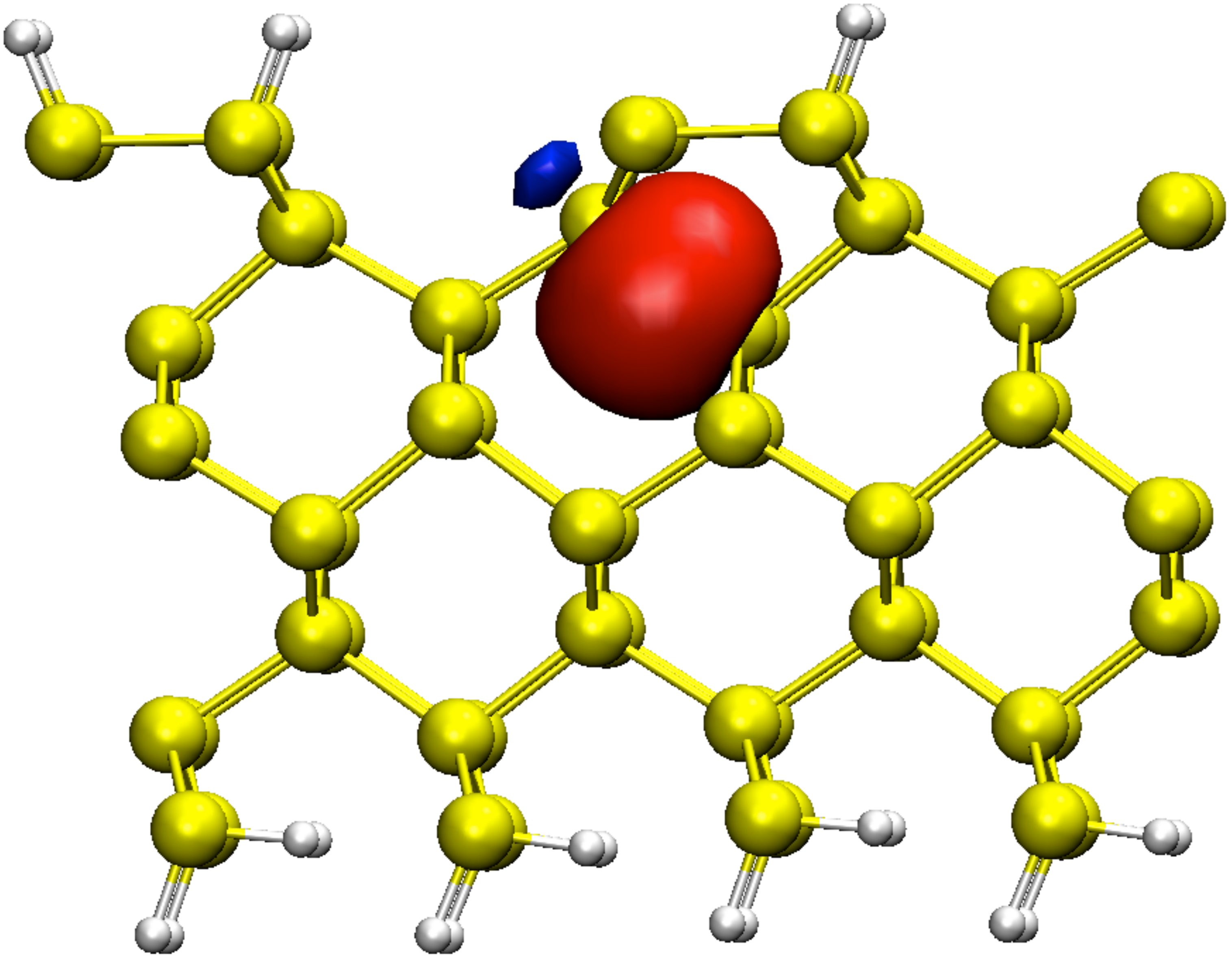}
	\end{center}
	\vspace{-0.5cm}
	\caption{Plot of a dangling bond-like (left), and a subsurface (right) Maximally
		Localized Wannier Functions. These MLWF are obtained by assigning
		one center per bond and allows one to recover the whole band structure
		of the system.}
	\label{fig:wannier}
\end{figure}

There is a large freedom in the choice of the Wannier functions used to describe the DB wire, 
in the sense that they are built to describe a certain energy interval of the
band structure of the DB wire, and, also, the points in space around which they
will be localized have to be specified as an input for the algorithm.
In a first step, we choose to assign one MLWF at each bond and one function per DB along the wire.
The chosen energy interval was $-12 < E - E_F < 3$ eV. A subset of the MLWFs
obtained for the ideal DB wire is depicted on Figure~\ref{fig:wannier}.

With the corresponding Hamiltonian (written in this MLWF basis set), we can study
the main electronic couplings in order to rationalize the electronic properties 
of the DB wire. One important feature is that the direct through space electronic coupling 
between nearest neighbour DBs is extremely low: $-0.06$ eV, as compared to the coupling between 
a DB and its subsurface functions that can reach $-1.70$ eV. 
Hence, one cannot think of transport as occurring only through direct hoping between
DBs  since the substrate dominates transport along the wire.

\begin{figure*}[!ht]
	\begin{center}
		\includegraphics*[scale=1]{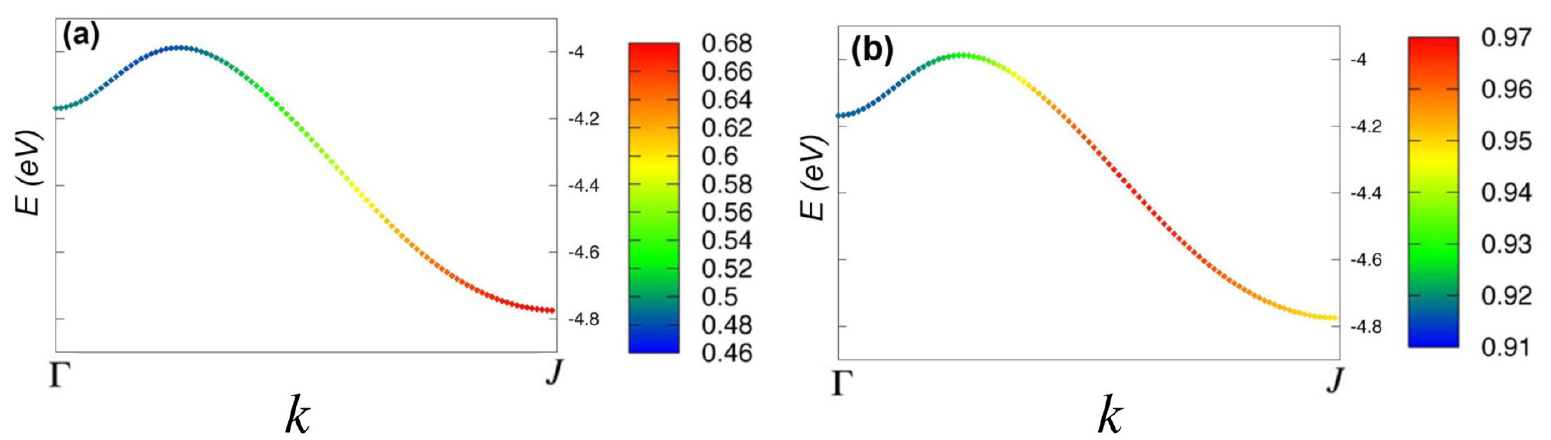}
	\end{center}
	\vspace{-0.5cm}
	\caption{Interpolated metallic band of the ideal wire along the direction of the
		DB wire. Color stands for the overlap of the eigenstate with MLWF
		corresponding to (a) the dangling-bonds, and (b) the dangling-bonds with
		8 subsurface functions (see Fig.~\ref{fig:wannier}). The Fermi level lies at
		$-4.25$ eV. The DB alone cannot account for the whole metallic band.
		Indeed, the band is well described only when taking into account 8 other functions.
		}
	\label{fig:proj}
\end{figure*}

From these functions, one can interpolate the band structure. The agreement with the
ab initio DFT-based band structure is excellent. As for the transport properties, we are
mainly interested in the metallic band that appears at the Fermi energy resulting
from the presence of the DBs. Projections of the metallic band on different sets of MLWFs
are depicted in Figure~\ref{fig:proj}. Here, the metallic band is correctly
described when including a large set of subsurface basis functions. This indicates that
the price to pay for using bond-like orbitals is that a larger number of them is required
in order to properly model the transport properties of non relaxed DBs chain, {\it i.e.} 
to describe the metallic band accurately.

\begin{figure}[!ht]
	\begin{center}
		\includegraphics*[height=3.5cm]{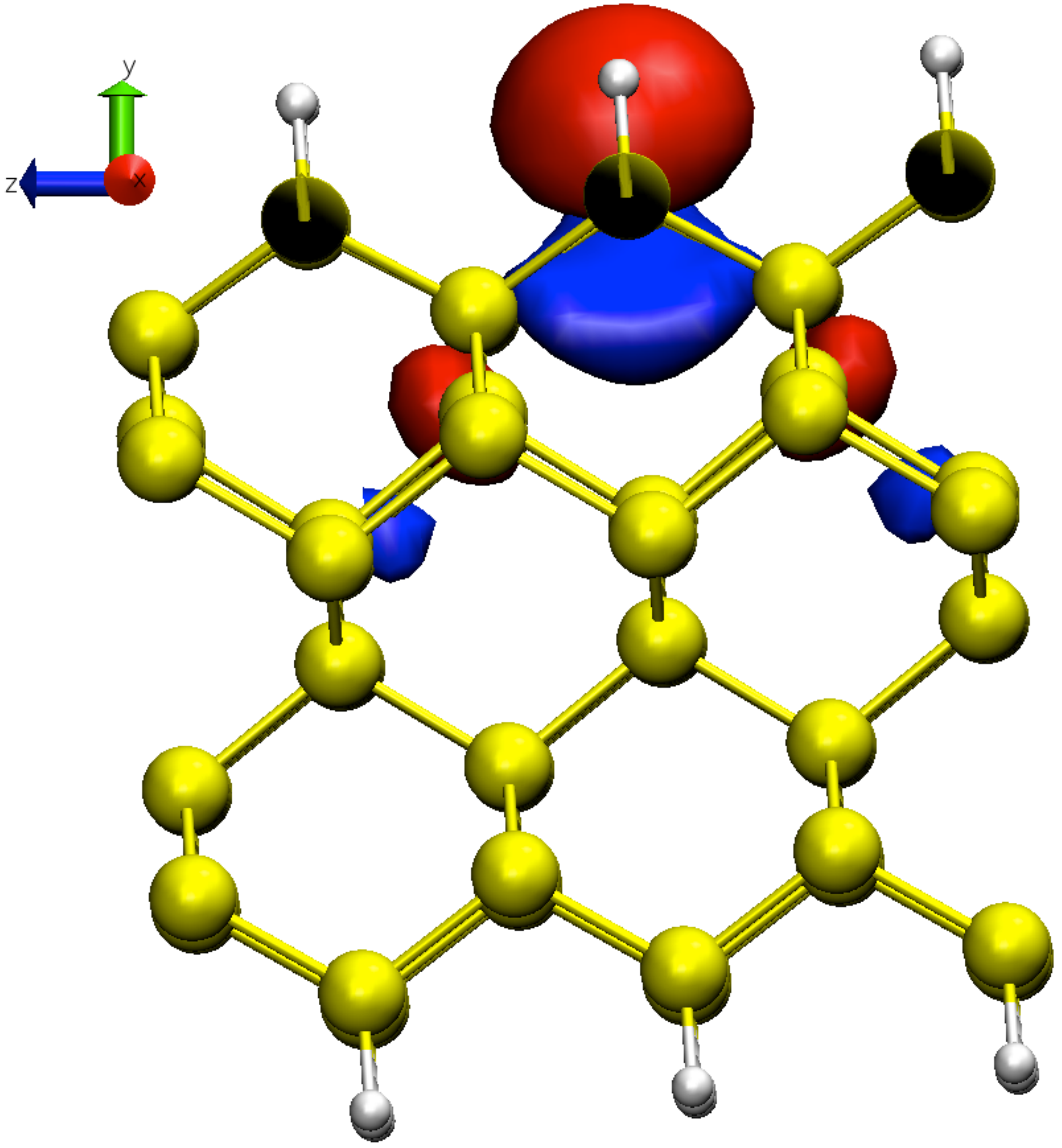}
		\hspace{.5cm}
		\includegraphics*[height=3.5cm]{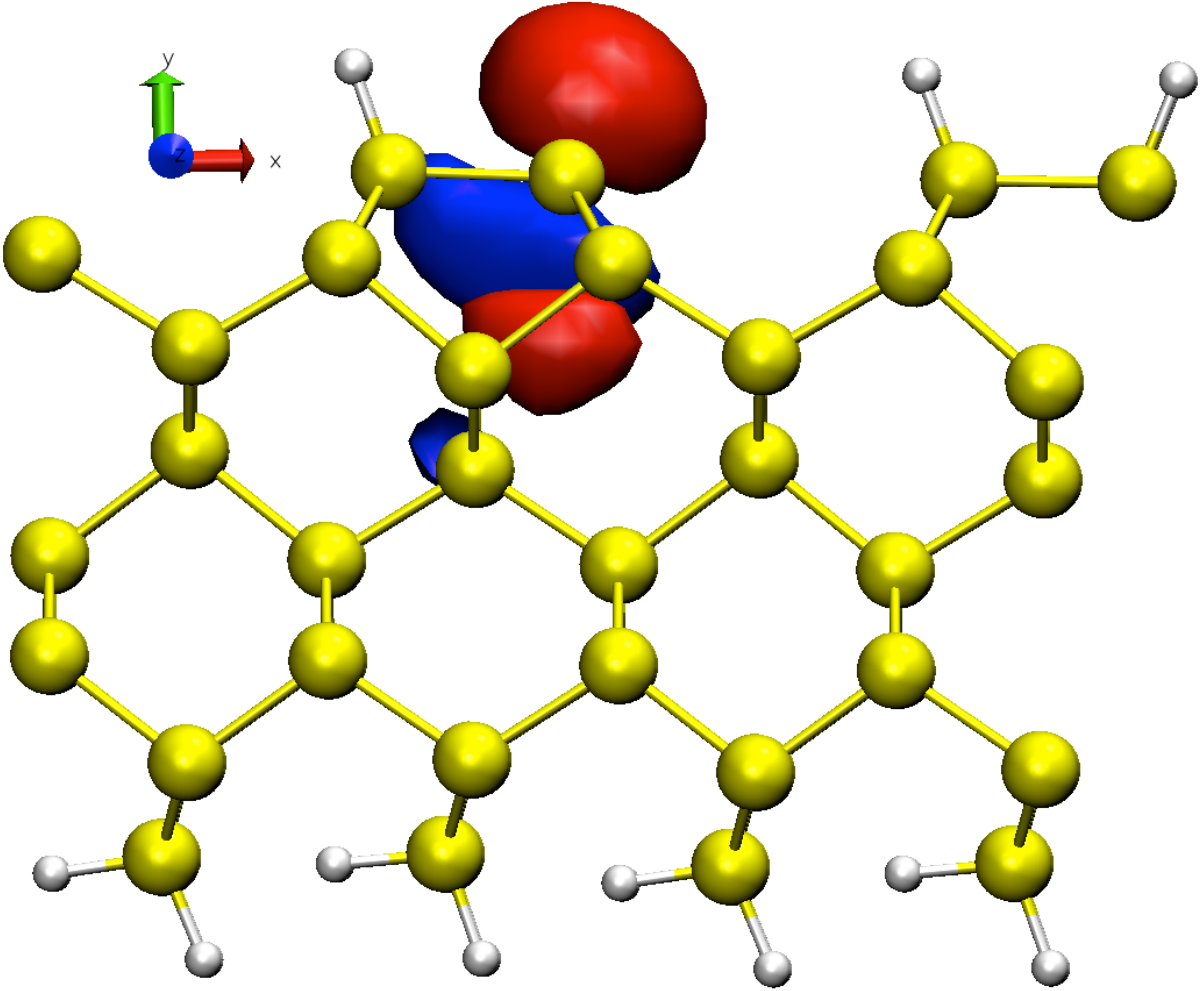}
	\end{center}
	\vspace{-0.5cm}
	\caption{Plot of the `effective' MLWF. This MLWF is obtained by assigning only one
		center per DB and by describing only the metallic band. It can be seen as a
		weighed `mix' between functions depicted in Figure~\ref{fig:wannier}.}
	\label{fig:wannier2}
\end{figure}

Alternatively, one can obtain a 1D tight-binding description of the system using only one
basis function by considering the MLWF using only the metallic band. This leads to the
calculation of an `effective' orbital (see Fig.~\ref{fig:wannier2}) that is also delocalized 
over the substrate.

The metallic band can then be computed from the MLWF Hamiltonian,
improving the calculations by including 
couplings to 2, 3 or more neighbors. We found that the coupling with
the fifth neighbor had to be included in the calculation in order to obtain a quantitative
agreement with the DFT-based result. In the next section, the model based on this
description of the wire will provide an interpretation of the scattering of electrons
on a single H-junction.

\section{Modeling the scattering on a single H-junction}
\label{sec:model}

As presented earlier (see Fig.~\ref{fig:transmission_ideal}), the single H-junction
has a transmission of $\sim 1$ in the energy window where it used to be
$T(E) = 2$ without any H atom. Furthermore, it quickly drops to values 
close to zero for lower energies. To model the presence of a saturated DB which acts 
as a scattering center, the scattering states~\cite{paulsson2007a, frederiksen2007a} will
be studied where the asymptotic states are the Bloch states of a periodic DB wire. 
We will compare the \textit{ab initio} transmission eigenchannels with the solutions 
found when modeling the DB wire as a one-dimensional chain with one orbital per DB site. 

\begin{figure}[!ht]
	\begin{center}
		\includegraphics*[scale=.28]{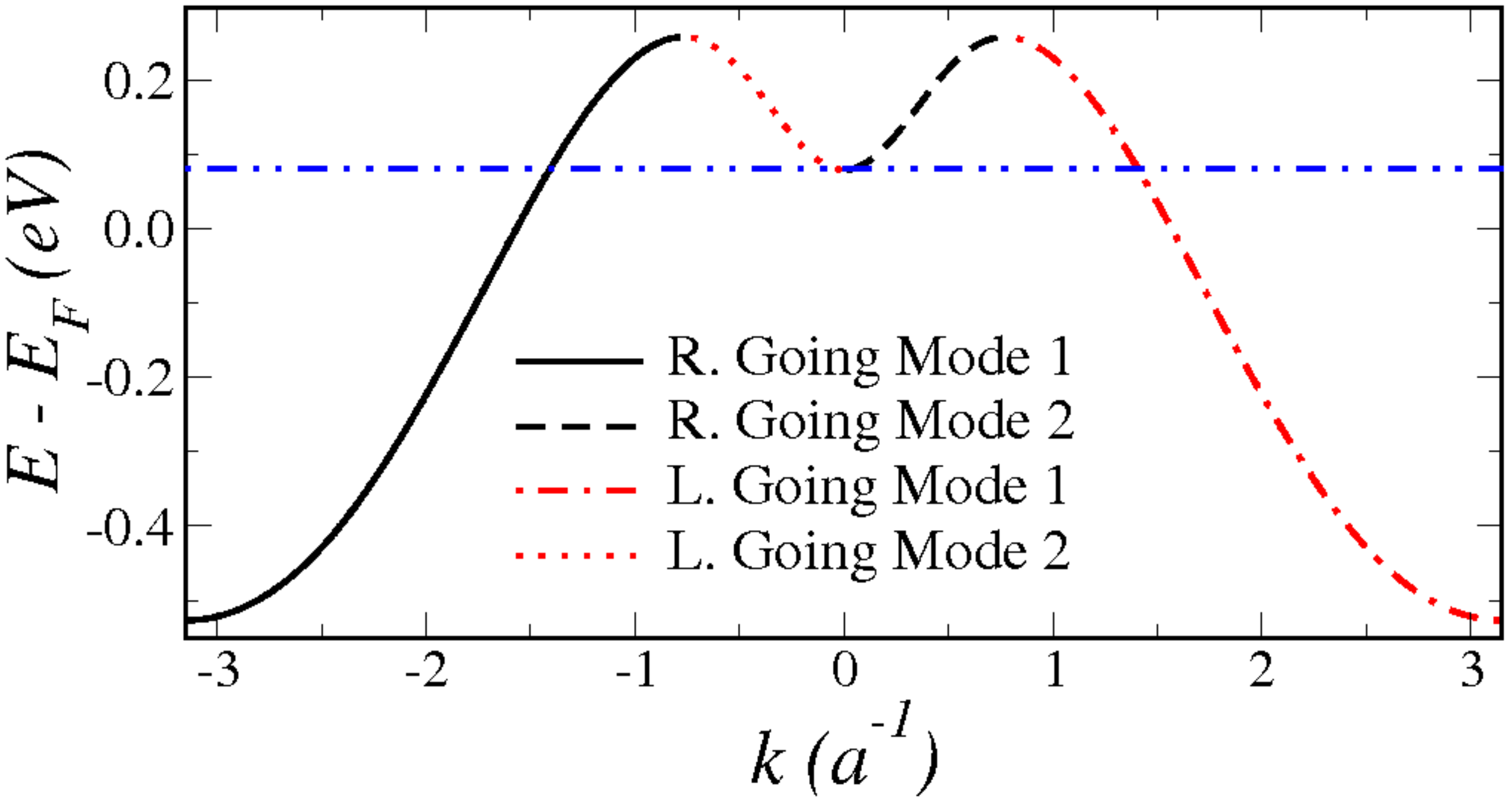}
	\end{center}
	\vspace{-0.5cm}
	\caption{Band structure of the ideal DB wire along the direction of the wire
		for positive and negative values of $k$. Right and left-going modes
		are indicated. Above the blue (dashed-double dotted line) there are two
		modes (channels) that give a maximum transmission of two, and bellow
		it only one. $a$ is the distance between two DBs.
	}
	\label{fig:as_modes}
\end{figure}

We start by characterizing the available asymptotic (Bloch) states at each energy. 
The Figure~\ref{fig:as_modes} shows the band structure with the corresponding right
and left-going states. They are labeled channel 1 and 2 respectively in the following.
It can be seen that depending on the energy
window, there are two or just one channel. When visualizing these states, the
change of the phase for the effective orbital, at each site, is determined by
the corresponding $k$ value of the channel.

\begin{figure}[!ht]
	\begin{center}
		\includegraphics*[scale=.16]{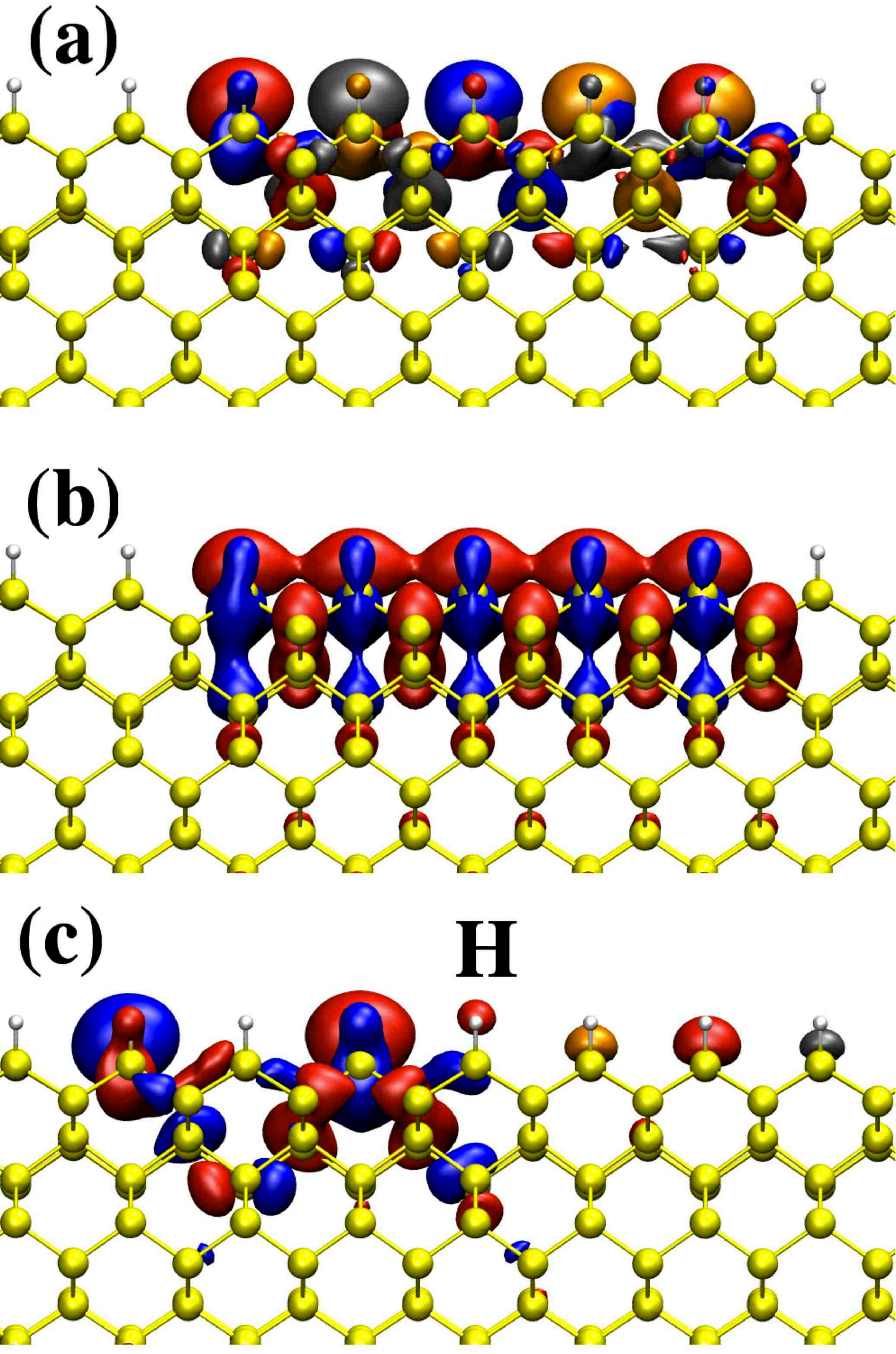}
	\end{center}
	\vspace{-0.5cm}
	\caption{(a) and (b) display the channels 1 and 2 at $E-E_F = 0.8$ eV, respectively.
		(c) displays the scattering state at $E-E_F = 0$.
		Color code: real, positive in blue; real, negative in red; imaginary, positive
		in silver; imaginary, negative in orange.}
	\label{fig:scatt_states}
\end{figure}

We move on now to the scattering states. For ideal wires without any scattering center, the
scattering states will correspond to the Bloch states at each energy (see
Fig.~\ref{fig:scatt_states}).
Both channels 1 and 2 show the expected phase modulation corresponding to $k$
vectors with values of $k_1 \sim -\pi /(2a)$ and $k_2 \sim 0$ (where $a$ is the distance
between two DBs).

\begin{figure}[!ht]
	\begin{center}
		\includegraphics*[scale=.35]{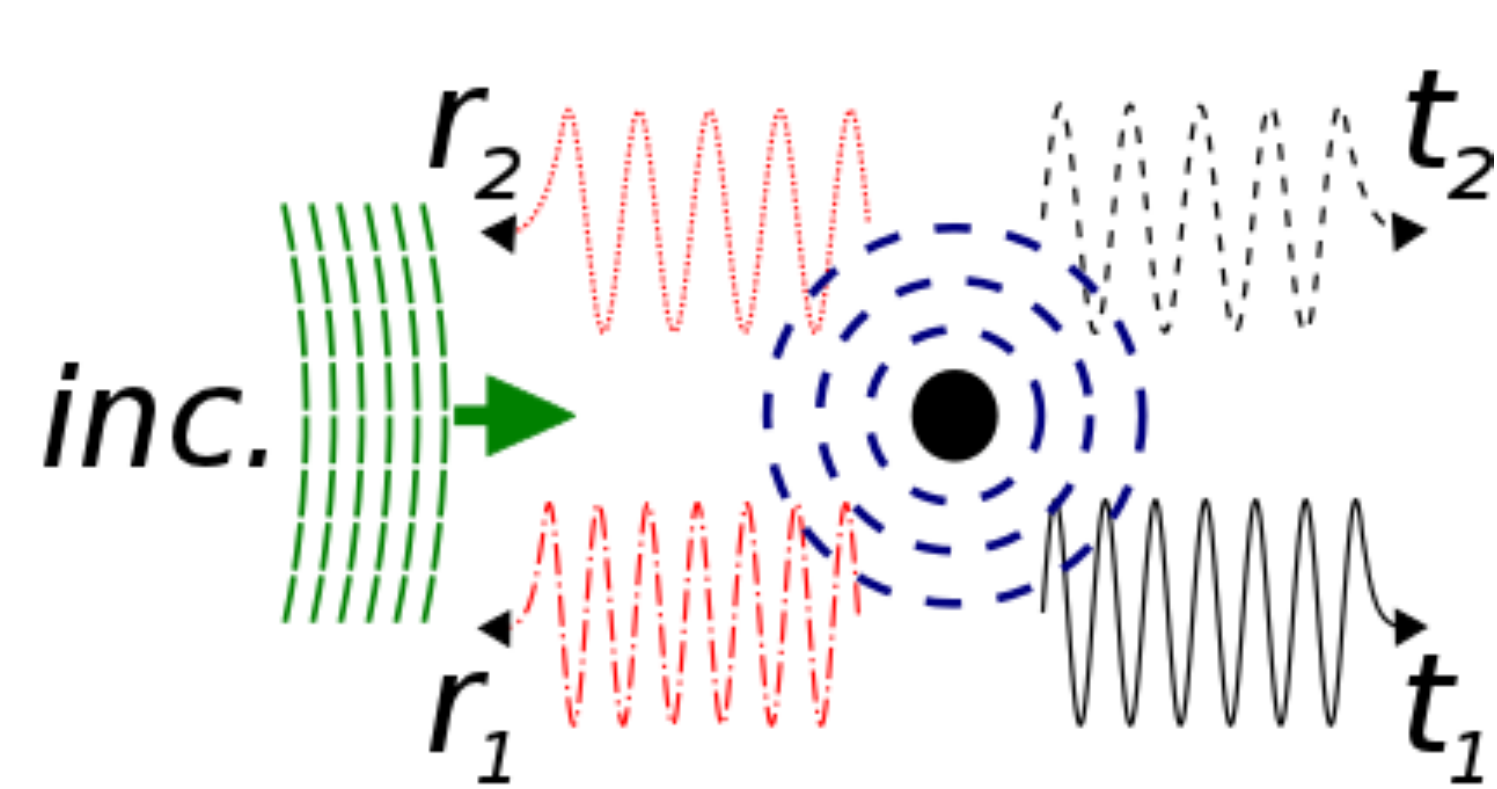}
	\end{center}
	\vspace{-0.5cm}
	\caption{Scattering of an incoming wave into the available outgoing
		modes (here for a case where there are two modes, both, to
		the left and to the right of the scatterer), weighed by the
		corresponding transmission and reflection coefficients.}
	\label{fig:mode_scatt}
\end{figure}

In the presence of an H atom (saturated DB), the incoming states will be scattered as shown
in Figure~\ref{fig:mode_scatt}. For energies values with just one channel and for a left
incoming state, the resulting scattering state will be a
superposition of the incoming wave (characterized by a certain $k_{inc}$) and the reflected
wave (with $k=-k_{inc}$) weighed by the (complex) coefficient $r$; and, to the right of the
scatterer, the transmitted wave will have $k=k_{inc}$, and will be weighed by the transmission
coefficient $t$.
As an example, in the case of the scattering state at $E-E_F = 0$ (see Fig.~\ref{fig:scatt_states} (c)),
for a right-going state coming from the left, $k_{inc} \sim -\pi /(2a)$ and $T \sim 0.1$. It can be seen
that the  transmitted wave corresponds to an outgoing wave with an associated $k \sim -\pi /(2a)$.
Since the transmission is low, with $r \sim 1$, the imaginary parts of the incoming and reflected
waves cancel, leaving only the real parts of the amplitudes.

The visualization process gets more complicated in the energy window where there are two
channels, since now the scattering can couple the two channels (see Fig.~\ref{fig:mode_scatt}).
Although the transmission and reflection coefficients could be explicitly obtained from
first-principles, we can now use the one-dimensional character of the problem and model the
system as a one dimensional chain, as explained above, using the parameters extracted from
the MLWF transformation. This strategy allows us to: (i) explicitly obtain the transmission and
reflection coefficients and hence the total transmission coefficient T(E) (that depends also on
the group velocities); (ii) obtain the transmission function in less than a second -- instead of
hours, the typical time it takes to obtain the T(E) from first-principles for the geometries
considered here. This latter aspect is particularly important if one wants to model more complex
arrangements involving more than one saturated DB in extended tunnel H-junctions.

\begin{figure}[!ht]
	\begin{center}
		\includegraphics*[scale=.32]{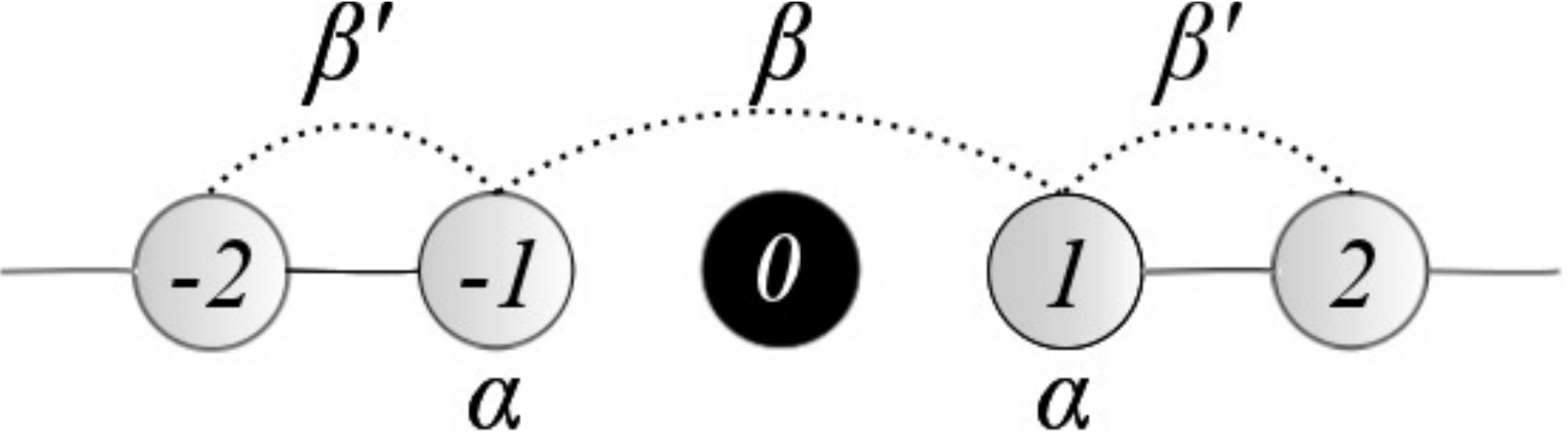}
	\end{center}
	\vspace{-0.5cm}
	\caption{One dimensional model for the DB wires. Site 0 represents a saturated DB, modeled
		as a defect, {\it i.e.} no transfer is possible through this site. With respect to the ideal DB
		wire, the Hamiltonian matrix element that is most affected is the one here represented
		by $\beta$, followed by $\alpha$ (onsite), and $\beta '$.}
	\label{fig:chain_1d}
\end{figure}

The saturated DB is modeled as a \emph{defect}, in the sense that, at the saturated site,
there is no \emph{available orbital for the electrons to be transfered through}, as depicted in
Figure~\ref{fig:chain_1d}. Different strategies can be used in order to solve the transport problem.
Here we have used the Multiband Quantum Transmitting Boundary Method,
proposed by Liang et al.~\cite{liang1998a} 

When modeling the changes in the tight-binding matrix elements due to the presence of the H atom,
one can expect that the most affected matrix elements are the electronic couplings between 
the two DBs separated by the saturated DB, in the sense of having their effective through
bond interaction reduced. Figure~\ref{fig:chain_1d} shows a scheme of the
one-dimensional model chain with one saturated DB (sites 1 and -1 are separated by a
saturated DB). The matrix elements indicated in Fig.~\ref{fig:chain_1d}, are obtained from a MLWF treatment
similar to the one applied on an ideal DB wire,
and given in Table~\ref{tab:model_params}. The corresponding transmission curve
is computed and shows an excellent agreement with the ab initio result. Also, for the energy window
where there are two channels, the decomposition of the transmission function in terms of the
contribution of the scattering of each channel into each other shows that they tend to be equally scattered.

The effect of the passivation of one DB is then to create a scattering defect that
effectively mixes the two incoming Bloch waves of different $k$ values.
This first inspection of the scattering properties leaves us with a physical picture for the electronic
transport in these systems. Bloch waves made up by the effective orbital, that are scattered by
a saturated DB as a defect in a linear chain. Due the long range electronic interactions through the substrate, this one-dimensional chain has a somehow unusual band structure, such that there can be two channels for certain energies that correspond to two Bloch states associated to two different $k$ values. 
This degeneracy gives rise to interesting scattering properties.

\begin{table}
	\caption{MLWF Hamiltonian matrix parameters that were changed when going from
		the ideal to a wire with one saturated DB (together with the introduction of a
		defect). All the values are given in eV and correspond to the transfer integral
		matrix element shown in Figure~\ref{fig:chain_1d}.}
	\label{tab:model_params}
	\begin{tabular}{cccc}
\hline
\hline
Chain type & $\beta$ & $\alpha$ & $\beta '$  \\
\hline
ideal & -0.053 & -4.333 & 0.190 \\
1H   & -0.041 & -4.298 & 0.182 \\
\hline
\hline
	\end{tabular}
\end{table}

\section{Dangling-bond wires}
\label{sec:dbwires}

As previously stated, an ideal finite DB wire is relaxing by performing either a Jahn-Teller distortion
(the NM wire) or a spin polarization with the DBs coupled ferro- or
antiferromagnetically (FM and AFM wires). The structures and the energetics of these different
atomic surface structures have been previously described.~\cite{lee2009a, nous2011a}
In the case of short DB-wires, the AFM state is the most stable. But the calculated
energy difference with other solutions remains too small to ignore them.~\cite{lee2009a, nous2011a}
Therefore, in this section, the transport  properties of the three NM, AFM and NM DB wires
configurations are discussed.

The system is divided into left and a right leads and a central scattering
region, in our case the different possible wires (Ideal, NM, AFM and
FM).~\cite{brandbyge2002a, datta_book} The self-consistent density matrix is converged
in the scattering region, using the open boundary conditions imposed by the leads
through their self energies, using the now standard Green's functions-based
method.~\cite{brandbyge2002a}
In order to avoid the sensitive problem of the description of the interface between realistic
electrodes and the DB-wires, we choose to take advantage of the metallic behavior of the
ideal wire (see section~\ref{sec:ideal}) and use them as the source for electrons to be
transferred through the scattering region. Thus, in our calculations, the role of right and
left leads are performed by semi-infinite ideal wires.

\begin{figure}[!ht]
	\begin{center}
		\includegraphics*[angle=180,scale=0.15]{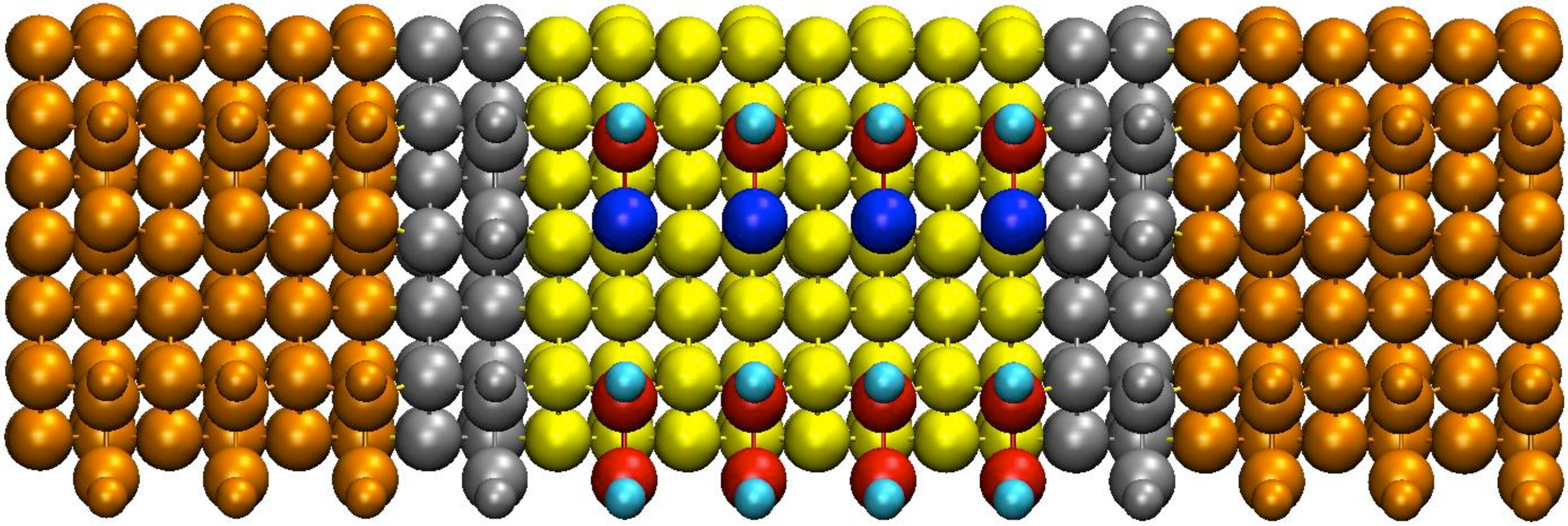}\\
	\end{center}
	\vspace{-0.3cm}
	\caption{Setup used for transport calculations. Semi-infinite ideal wires (orange) are
		used as electrodes. The DB-wire is decoupled from the electrodes by
		H-passivated dimers (gray).}
	\label{fig:fw}
\end{figure}
Notice that the central scattering region include (i) a finite DB-wire with a length going from 2 to
5 DBs and (ii) two H-passivated dimers placed at each end of this DB wire (see Fig.~\ref{fig:fw}).
Those H-junctions are used here to decouple the central DB wire from the leads. As a result,
one can converge the different solutions corresponding to the different states of the DB-wire,
while the leads remain as ideal DB wires.

\begin{figure}[!ht]
	\begin{center}
		\includegraphics*[scale=0.55]{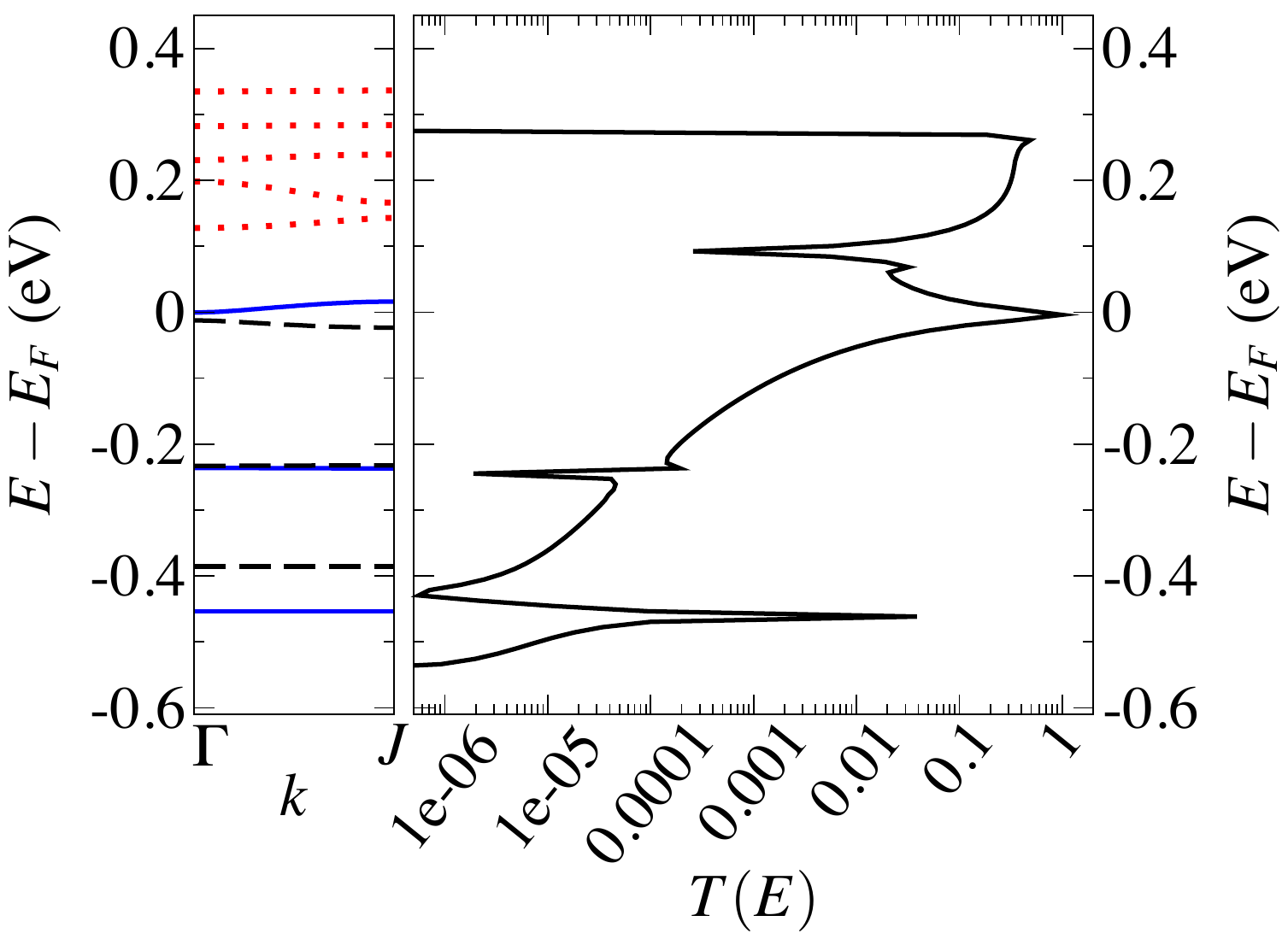}\\
	\end{center}
	\vspace{-0.5cm}
	\caption{Band structure (left) and transmission $T(E)$ (right) of the 5-DB ideal wire.
		Bands in blue (solid lines) correspond to states purely localized on the wire,
		bands in black (dashed lines) to those localized on the electrodes. The bands
		in red (dotted lines) result from a mixing between the states of the electrodes 
                and those of the wire.
		The bands of the electrodes appear shifted due to the periodic boundary
		conditions used to calculate the band structure.}
	\label{fig:resonance}	
\end{figure}
As one can then expect, the effect of the two H-passivated dimers is to lead to a
confinement of the DB-wire states, thus a particle in a box-like behavior.
If instead of using the open boundary conditions, one uses periodic conditions, these
confined states will appear as essentially flat bands within the bulk gap as can be
seen on Figure~\ref{fig:resonance} on the example of a 5 DB ideal wire (blue solid lines).
The corresponding transmission exhibits peaks and dips that can be understood by
examining the confined states.

\begin{figure}[!ht]
	\begin{center}
		\includegraphics*[scale=0.55]{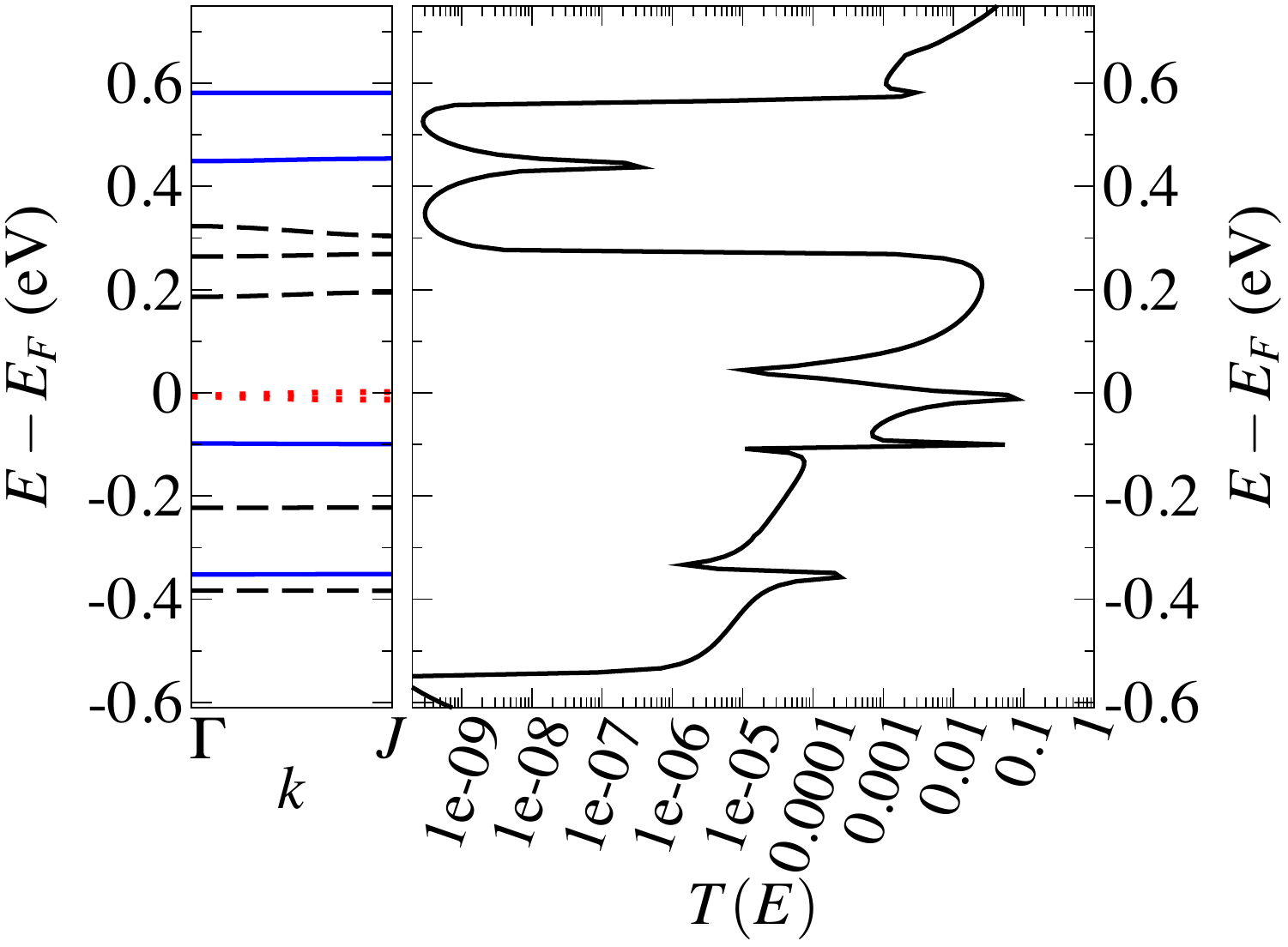}\\
	\end{center}
	\vspace{-0.5cm}
	\caption{Same as Fig.~\ref{fig:resonance} for the 5-DB NM wire.}
	\label{fig:nm5db}	
\end{figure}

The three first states of an ideal 5-DB wire give clear resonance peaks in the transmission
spectrum. They are within the energy range of the one-channel energy range of the ideal
DB wire lead, {\it i.e.} for ($-0.53$ eV $< E~-~E_F < 0.08$ eV). On the other hand, the
states located in the two-channel energy range result from a mixing between the states of
the DB wire and the ones from the leads. As a result, the transmission spectrum does not
display resonance peaks but an overall enhancement in the high energy range of the
two-channel band.

\begin{figure}[!ht]
	\begin{center}
		\includegraphics*[scale=0.2]{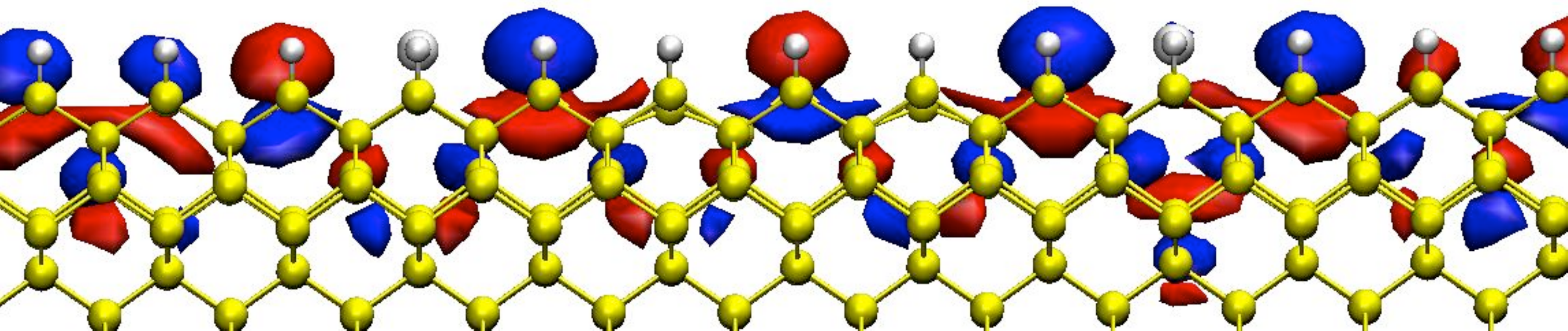}\\
	\end{center}
	\vspace{-0.5cm}
	\caption{Plot of wavefunction displaying contributions on the wire and on the electrodes
		in the 5-DB NM wire (red dotted lines on Fig.~\ref{fig:nm5db}) at $\Gamma$.}
	\label{fig:nmwf}
\end{figure}

The T(E) spectral behavior is similar for a NM wire (see Fig.~\ref{fig:nm5db}). In the case a 5-DB
NM wire, the DB wire's states give also rise to T(E) resonances. The larger transmission is obtained
for a DB-wire state coupled with a state of the leads (see Fig.~\ref{fig:nmwf}).
In the two-channel energy range, the T(E) spectrum is the one observed in the case of H-junctions
(see Fig.~\ref{fig:transmission_ideal}). Indeed, as opposed to the 5-DB ideal wire, the
transmission is not perturbed by wire's states that are shifted and give resonance peaks
at higher energies. When going from a 2 to a 5-DB NM wire, the overall shape of the transmission 
\begin{figure}[!ht]
        \begin{center}
                \includegraphics*[scale=0.65]{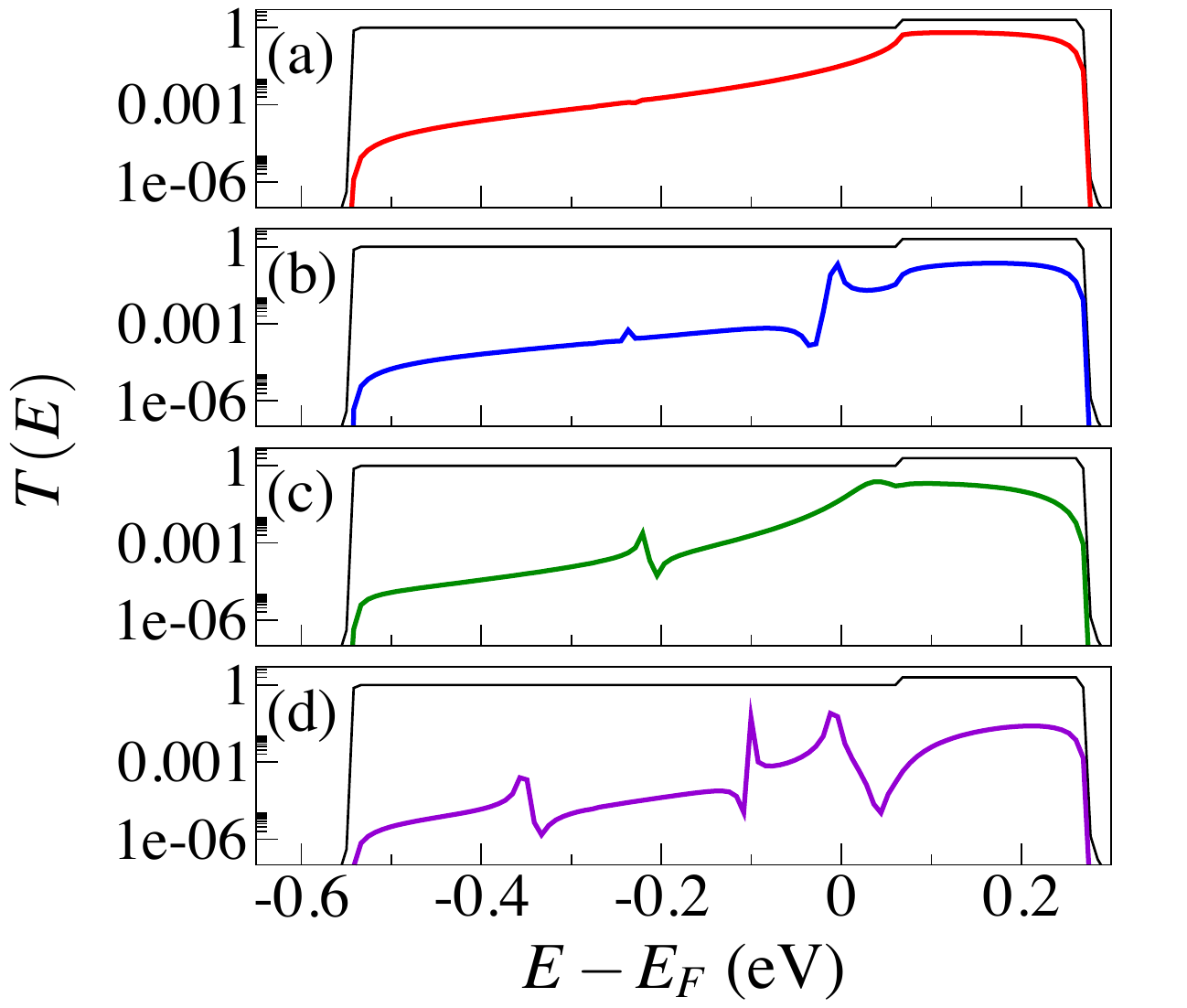}
        \end{center}
        \vspace{-0.5cm}
        \caption{Transmission as a function of energy through NM wires. 2, 3, 4, and 5-DB wires
                are depicted in (a), (b), (c) and (d), respectively. The transmission of the
                ideal wire is depicted in black.}
        \label{fig:nm}
\end{figure}
spectrum remains the same but with a larger number of resonance peaks corresponding to an increasing
number of DB in the wire (see Fig.~\ref{fig:nm}).~\footnote{Let us note that some
resonance peaks are pushed outside of the energy range defined by the leads and, thus, do not
show up on Figure~\ref{fig:nm} T(E) spectra.}

\begin{figure}[!th]
	\begin{center}
		\includegraphics*[scale=0.55]{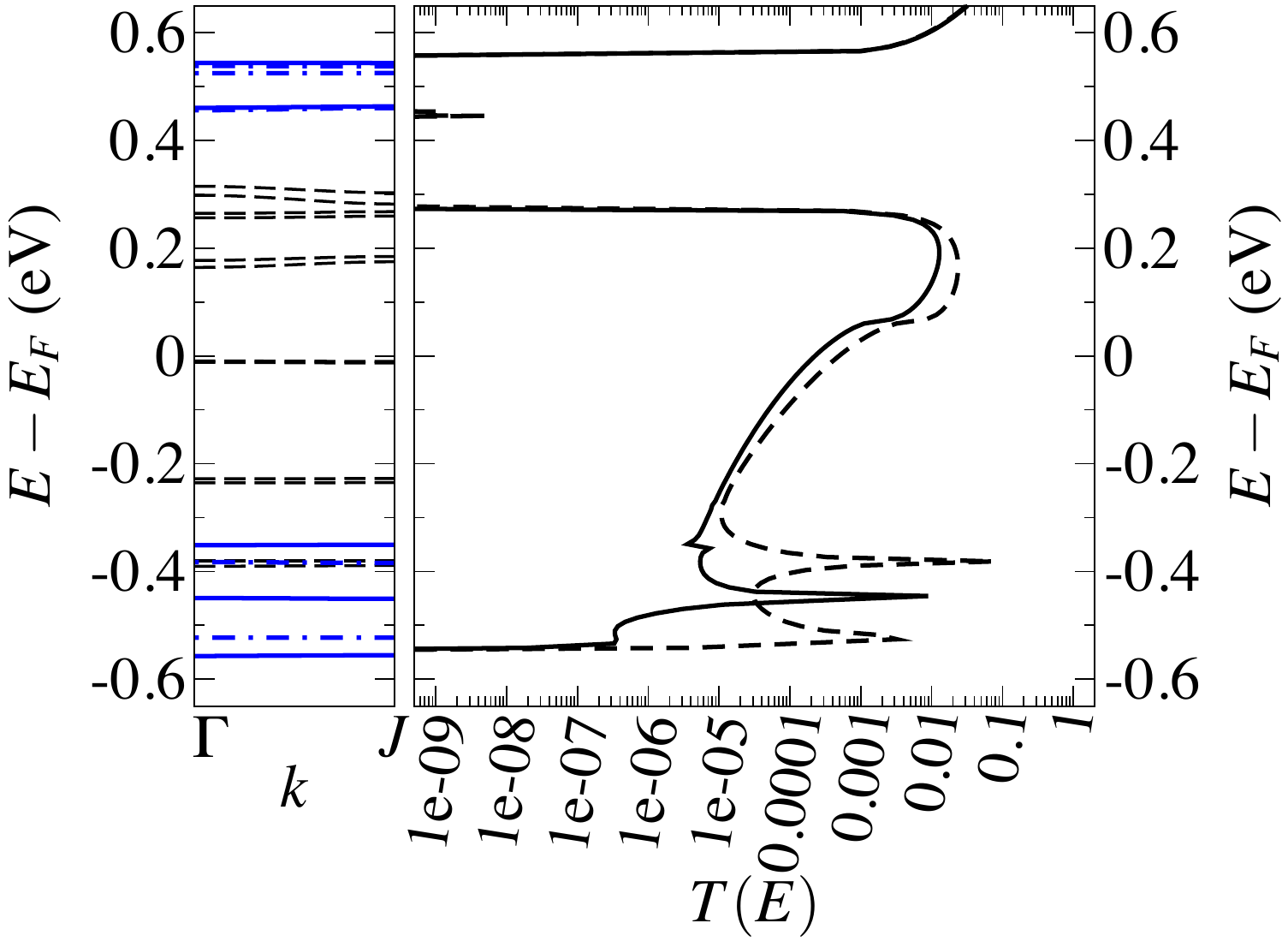}\\
	\end{center}
	\vspace{-0.5cm}
	\caption{Band structure (left) and transmission $T(E)$ (right) of the 5-DB AFM wire.
		The states localized on the wire are depicted in blue (solid and dashed lines for
		majority and minority spins, respectively). States of the electrodes are depicted
		in black dashed lines. The bands of the electrodes appear shifted due to the
		periodic boundary conditions used to calculate the band structures.}
	\label{fig:afm5db}
\end{figure}

\begin{figure}[!ht]
	\begin{center}
		\includegraphics*[scale=0.65]{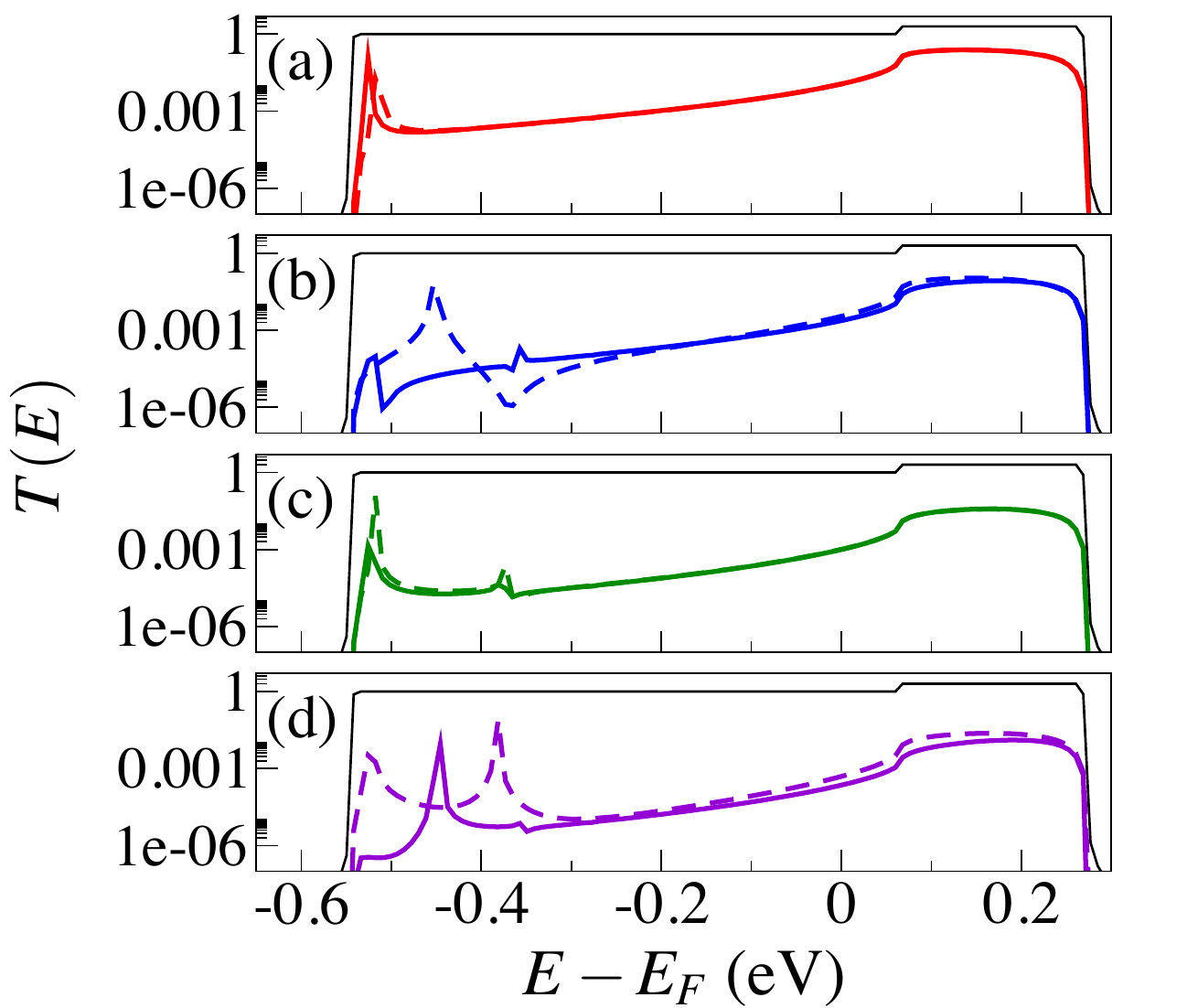}\\
	\end{center}
	\vspace{-0.5cm}
	\caption{Transmission through AFM wires as a function of energy. (a) 2-DB length, (b) 3 DB, (c) 4 DB and (d) 5 DB.
		Majority and minority spins are depicted in solid and dashed line,
		respectively. The transmission of the ideal wire is depicted in black.
                }
	\label{fig:afm}
\end{figure}

In the same manner, the transmission through a 5-DB AFM wire exhibits resonances
corresponding to states localized within the wire (see Fig.~\ref{fig:afm5db}). Nevertheless,
these states are confined below $E - E_F = -0.3$ eV and above 0.3 eV, leaving the
transmission largely undisturbed. The situation is the same whatever the length of the
wire (see Fig.~\ref{fig:afm}).
Let us stress that the difference between the transmission of both spins states in the case of
odd finite length DB wires is due to the monoreferencial nature of the DFT approach that 
cannot deliver proper spin states.
We observe an exponential decay of the transmission away from the resonances.
Thus, one can define and evaluate, as in the H-junction case, an inverse decay length
for the AFM wire. The calculated value $\gamma_{AFM} = 0.89$~\AA$^{-1}$ at
$E - E_F = 0.15$ eV is over three times larger than the one of the H-junctions at the
same energy.

\begin{figure}[!ht]
	\begin{center}
		\includegraphics*[scale=0.55]{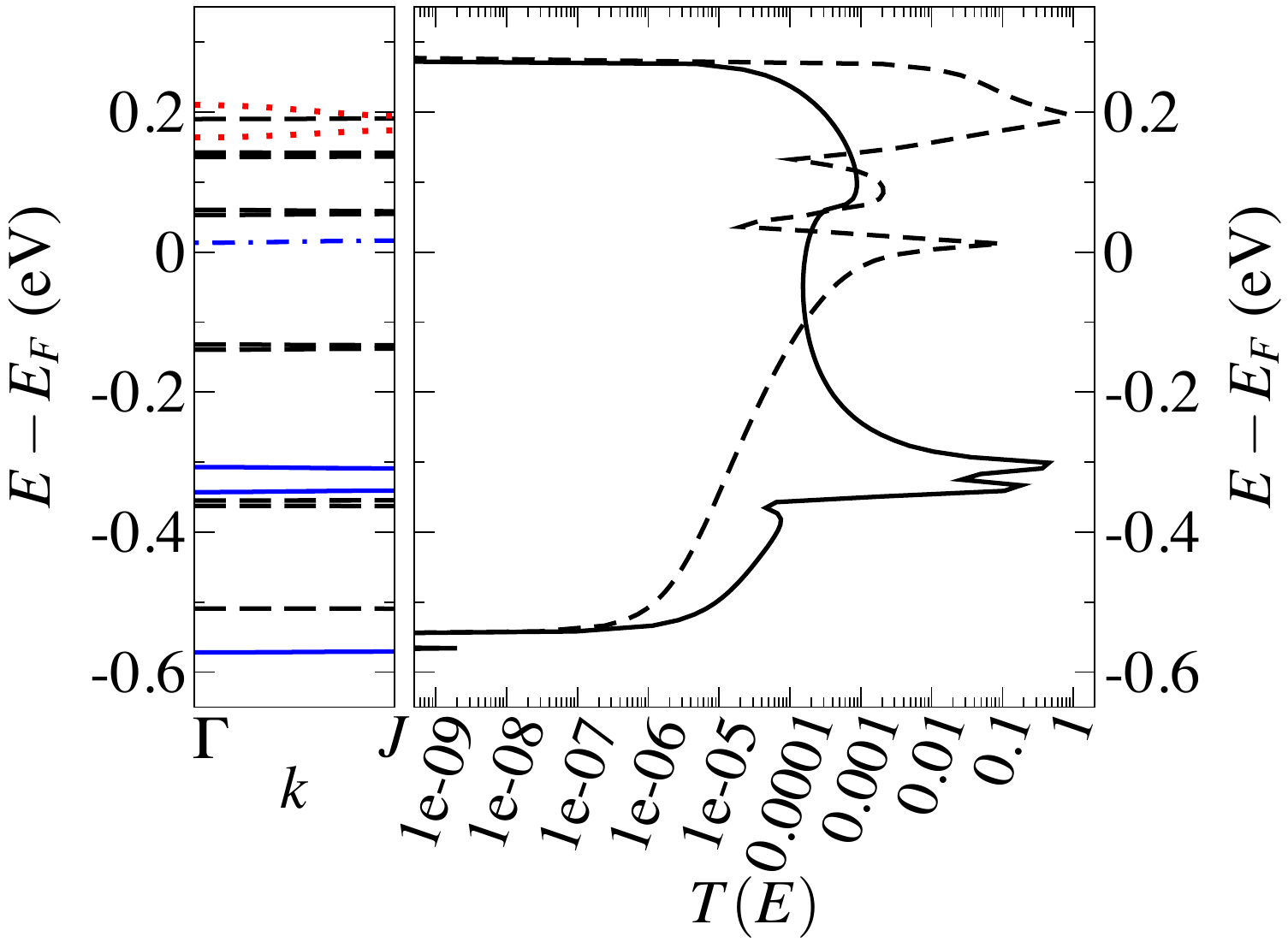}\\
	\end{center}
	\vspace{-0.5cm}
	\caption{Band structure (left) and transmission $T(E)$ (right) of the 5-DB FM wire.
		The states localized purely on the wire are depicted in blue (solid and dashed
		line for majority and minority spins, respectively). States of the electrodes are
		depicted in black dashed lines. Bands in red dotted lines result from a mixing
		between electrodes' and wire's states for the minority spin. The bands of the
		electrodes appear shifted due to the periodic boundary conditions used to
		calculate the band structures.}
	\label{fig:fm5db}	
\end{figure}

\begin{figure}[!ht]
	\begin{center}
		\includegraphics*[scale=0.65]{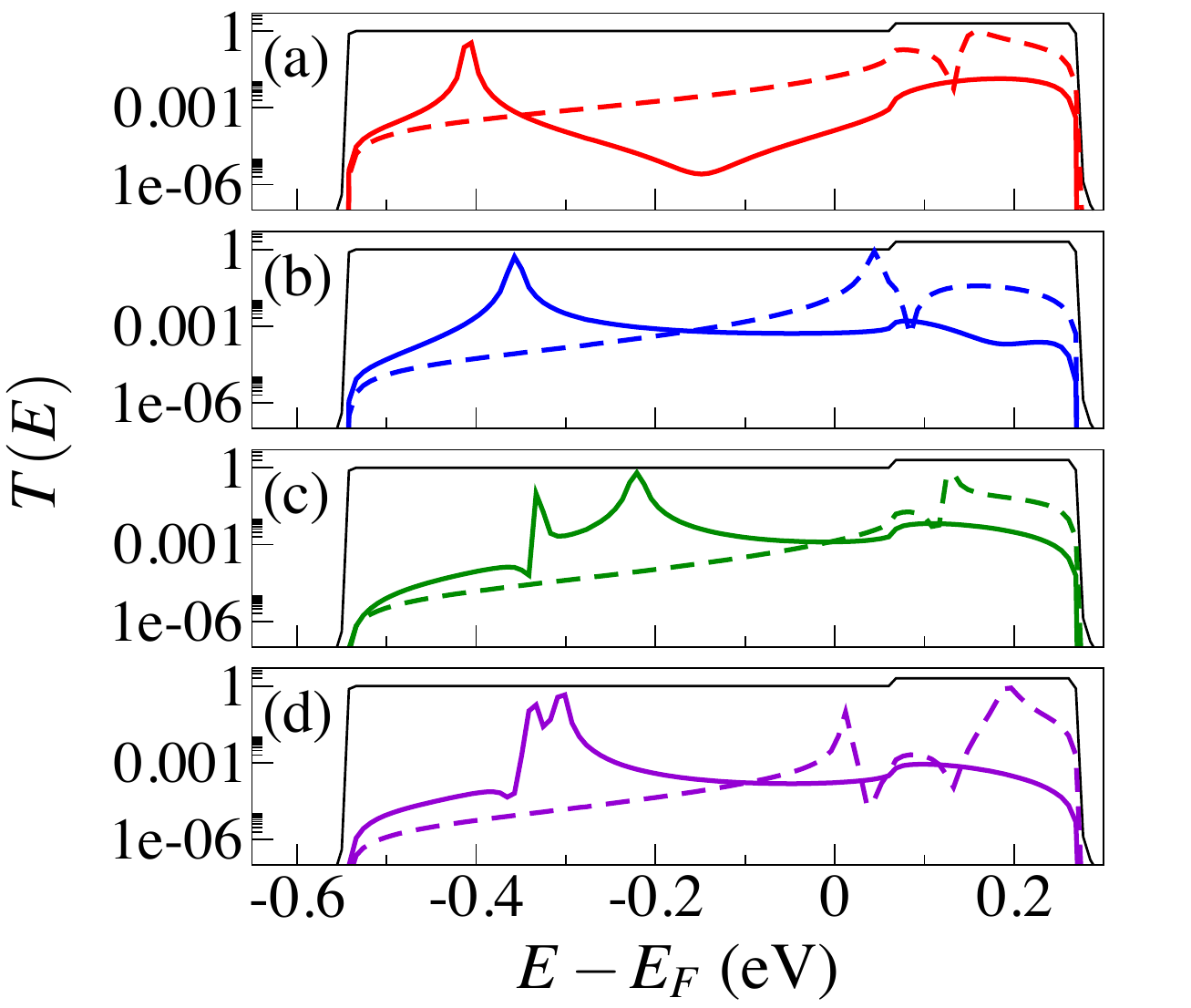}\\
	\end{center}
	\vspace{-0.5cm}
	\caption{Transmission through FM wires as a function of energy. 
The different pannels and lines correspond to the same cases of Fig.~\ref{fig:afm}.
                }
	\label{fig:fm}
\end{figure}
The case of the 5-DB FM wire is extremely similar to the previous one (see Fig.~\ref{fig:fm5db}).
The main difference lies in the energy distribution of the DB wire's states depending on their
spin. Indeed, as expected, the majority spin states occupy a lower range of energy, whereas
the minority spin states are found near the two-channel energy range.
As a consequence, the transmission spectra are extremely different for both spins, whatever the length
of the finite central DB wire (see Fig.~\ref{fig:fm}). The energy localization of the resonances is proper
to the spin, featuring a spin filter behavior.

This systematic inspection reveals  a common feature in the transmission of finite size
DB wires due to an overall general behavior of $T(E)$. It is due to
the original shape of the ideal DB wire, used here as electrode. This smooth
transmission function is then perturbed by multiple resonances arising from states
of the wire confined in between passivated dimers. The positions of these peaks and
dips is specific to the nature of the wire as we have just seen.

Figures~\ref{fig:nm}, \ref{fig:afm} and \ref{fig:fm} show the electron transmission coming
from an ideal DB wire, transmitting into another ideal DB wire at a given electron energy
$E$. As such, it is not easy to deduce the actual electron current in the studied systems.
We have then computed the electron current using the Landauer-B{\"u}ttiker
formula:~\cite{datta_book}
\begin{equation}
I = \frac{2 e}{h} \int_{-\infty}^{\infty} T(E,V) [ f_R(E)-f_L(E)] dE.
\end{equation}
Where, $T(E,V)$ is the transmission function for an electron of energy $E$ when the bias
between the two DB electrodes is $V$, and $f_R(E)$ ($f_L(E)$) is the right- (left-)
electrode Fermi occupation function. We further simplify the current $I$ calculation using
the zero-bias tansmissions of Fig.~\ref{fig:nm}, \ref{fig:afm} and \ref{fig:fm}.
\begin{figure}[htbp]
        \begin{center}
                \includegraphics*[scale=0.68]{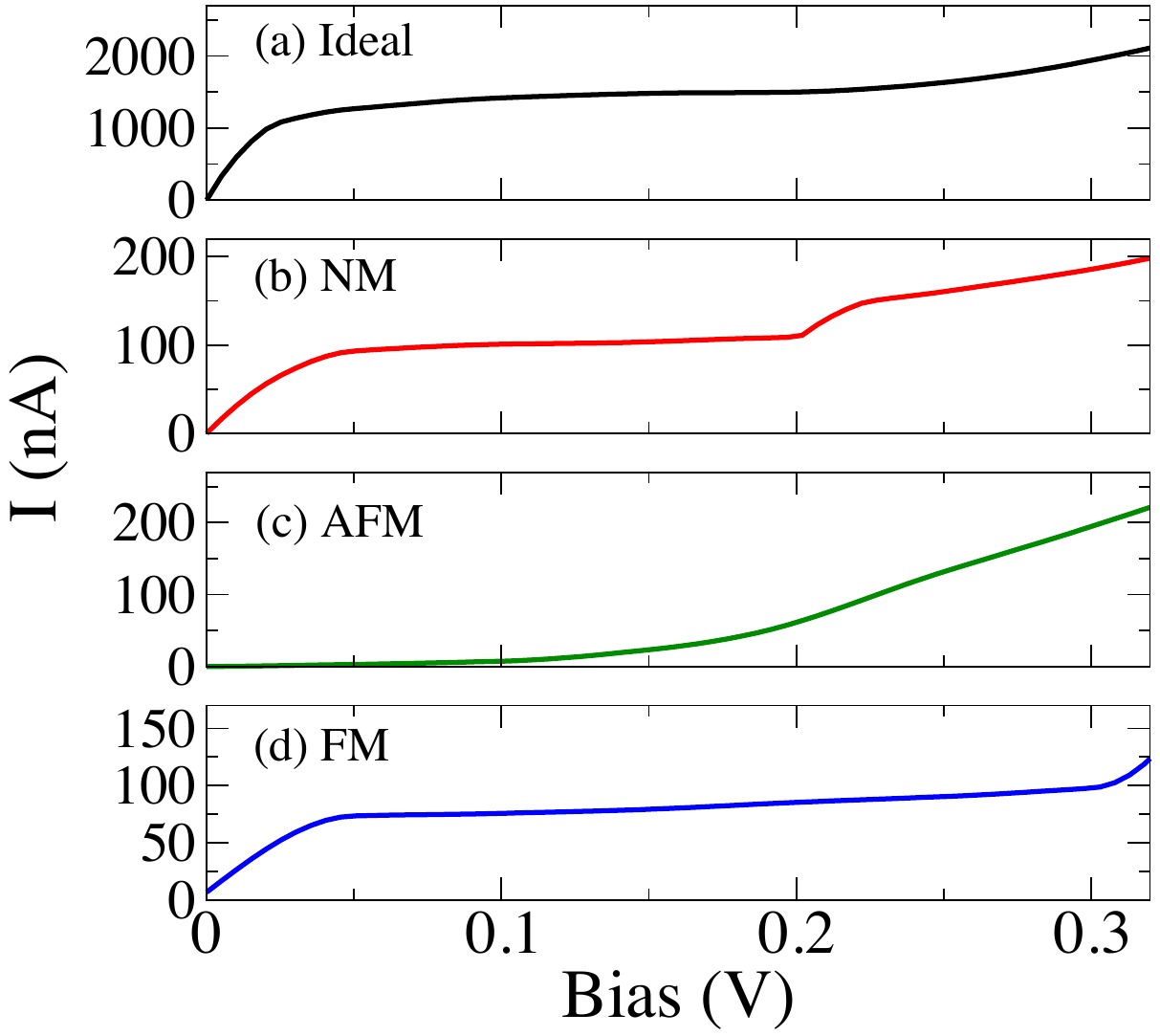}\\
        \end{center}
        \vspace{-0.5cm}
        \caption{Calculated current (nA) with respect to the bias (V) for 5-DB wires.
                Ideal, NM, AFM and FM wires are displayed top to bottom.}
        \label{fig:current}
\end{figure}
Figure~\ref{fig:current} shows the computed I-V curves for 5-DB wires in the ideal,
NM, AFM and FM configuration. As we can see, the electronic currents of actual
physical wires are very reduced as compared to the current of the ideally
undistorted non-magnetic  wire.

Our recent study of total energy and stability of DB wires~\cite{nous2011a} shows
that the AFM and NM solutions are thermodynamically coexisting, however their
very different I-V characteristics would permit to identify the type of obtained DB wire.

\section{Polaronic effects in transport}

The above study only considers elastic transport. However,  the large
electron-vibration coupling leading the DB wires to Jahn-Teller distortions
would modify transport from our above description.
Indeed, a DFT-based tight-binding study suggested that transport in
infinite DB wires would take place in the form of small-polaron
diffusion.~\cite{bowler2001a,todorovic2002a}
Hole polarons are extended over 3-DB sites,~\cite{bowler2001a} while
electron polarons present a similar confinement.~\cite{bowler2001a}
We have estimated the extension of the induced electron polaron, by including
an extra charge in the NM 5-DB wire. The most noticeable effect is
that the distortion reverts sign: DB sites moving away from the surface when
before they were moving against the surface and viceversa. But the degree of distortion
remains the same. More importantly, the electron is not localized to only
three sites, but it extends over the full wire, giving rise to no confinement except
for the finite size of the DB wire. Hence, the existence of small
polarons may be due to the actual extension of the DB wire implying that transport will
take place by polaron hopping only in long wires.~\cite{todorovic2002a}

The study of the electronic ground state of the negatively charged 5-DB wire further
shows that there is little difference in the electronic states of the neutral and charged
wires. Indeed, one can recover the other wire's electronic structure just by shifting the
Fermi level. In this way, the new {\em Highest Occupied Molecular Orbital} (HOMO)
is basically the former  {\em Lowest Unoccupied Molecular Orbital} (LUMO) of the
neutral 5-DB wire. Therefore, despite the need of including electron-vibration
coupling in the quantitative description of DB wire transport,~\cite{futuro} our present
results are qualitative transport descriptions because the electonic current is largely
tunneled via electronic states closely resembling the neutral DB-wire states.

\section{Conclusion}
In this study, we investigated the electronic
transport properties of dangling-bond (DB) silicon wires
on H passivated Si(100). Thanks to DFT calculations and 
its analysis by maximally localized Wannier functions,
we have been able to rationalize the 
transport properties of these DB wires. As an example, we have
shown the effect of a Hydrogen impurity as a source of electron
scattering in the wire. 
This study shows that transport mainly proceeds via subsurface atoms
because the direct DB interaction is negligible.
A Hydrogen impurity is then efficient in interrupting the subsurface
transport because it decouples the subsurface
states from the surrounding DB. In this
way, a Hydrogen impurity mixes and scatters the channels of the
otherwise unperturbed wire.

Different  finite-size dangling-bond wires have been studied.
We have considered
unperturbed, Jahn-Teller distorted, antiferromagnetic
and ferromagnetic wires containing 2, 3, 4 and 5 DB's. Each wire displays
typical and unique trends in the transmission, allowing us
 to characterize the nature of the
wire from its transport properties. The size of the studied wires
are in the order of the small-polaron extension of DB wires,~\cite{bowler2001a} 
producing a shifting
of the electronic structure but qualitative similar features. Hence, we expect
polaronic effects mainly to be important in long wires.
The transport properties revealed in this study will permit to characterize
the DB wires experimentally accessible.~\cite{hitosugi1999a}


\section*{Acknowledgments}
This work has been supported by the European Union Integrated Project AtMol (http://www.atmol.eu).
We acknowledge CESGA for computational resources.
The authors thank F. Ortmann for fruitful discussions.

\bibliography{kepenekian_et_al}

\begin{thebibliography}{10}%
\makeatletter
\providecommand \@ifxundefined [1]{%
 \ifx #1\undefined \expandafter \@firstoftwo
 \else \expandafter \@secondoftwo
\fi
}%
\providecommand \@ifnum [1]{%
 \ifnum #1\expandafter \@firstoftwo
 \else \expandafter \@secondoftwo
\fi
}%
\providecommand \enquote [1]{``#1''}%
\providecommand \bibnamefont  [1]{#1}%
\providecommand \bibfnamefont [1]{#1}%
\providecommand \citenamefont [1]{#1}%
\providecommand\href[0]{\@sanitize\@href}%
\providecommand\@href[1]{\endgroup\@@startlink{#1}\endgroup\@@href}%
\providecommand\@@href[1]{#1\@@endlink}%
\providecommand \@sanitize [0]{\begingroup\catcode`\&12\catcode`\#12\relax}%
\@ifxundefined \pdfoutput {\@firstoftwo}{%
 \@ifnum{\z@=\pdfoutput}{\@firstoftwo}{\@secondoftwo}%
}{%
 \providecommand\@@startlink[1]{\leavevmode\special{html:<a href="#1">}}%
 \providecommand\@@endlink[0]{\special{html:</a>}}%
}{%
 \providecommand\@@startlink[1]{%
  \leavevmode
  \pdfstartlink
   attr{/Border[0 0 1 ]/H/I/C[0 1 1]}%
   user{/Subtype/Link/A<</Type/Action/S/URI/URI(#1)>>}%
  \relax
 }%
 \providecommand\@@endlink[0]{\pdfendlink}%
}%
\providecommand \url  [0]{\begingroup\@sanitize \@url }%
\providecommand \@url [1]{\endgroup\@href {#1}{\urlprefix}}%
\providecommand \urlprefix [0]{URL }%
\providecommand \Eprint[0]{\href }%
\@ifxundefined \urlstyle {%
  \providecommand \doi [1]{doi:\discretionary{}{}{}#1}%
}{%
  \providecommand \doi [0]{doi:\discretionary{}{}{}\begingroup
  \urlstyle{rm}\Url }%
}%
\providecommand \doibase [0]{http://dx.doi.org/}%
\providecommand \Doi[1]{\href{\doibase#1}}%
\providecommand \bibAnnote [3]{%
  \BibitemShut{#1}%
  \begin{quotation}\noindent
    \textsc{Key:}\ #2\\\textsc{Annotation:}\ #3%
  \end{quotation}%
}%
\providecommand \bibAnnoteFile [2]{%
  \IfFileExists{#2}{\bibAnnote {#1} {#2} {\input{#2}}}{}%
}%
\providecommand \typeout [0]{\immediate \write \m@ne }%
\providecommand \selectlanguage [0]{\@gobble}%
\providecommand \bibinfo [0]{\@secondoftwo}%
\providecommand \bibfield [0]{\@secondoftwo}%
\providecommand \translation [1]{[#1]}%
\providecommand \BibitemOpen[0]{}%
\providecommand \bibitemStop [0]{}%
\providecommand \bibitemNoStop [0]{.\EOS\space}%
\providecommand \EOS [0]{\spacefactor3000\relax}%
\providecommand \BibitemShut [1]{\csname bibitem#1\endcsname}%
\bibitem{muller1999a}%
  \BibitemOpen
  \bibfield{author}{%
  \bibinfo {author} {\bibfnamefont{D.~A.}\ \bibnamefont{Muller}}, \bibinfo
  {author} {\bibfnamefont{T.}~\bibnamefont{Sorsch}}, \bibinfo {author}
  {\bibfnamefont{S.}~\bibnamefont{Moccio}}, \bibinfo {author}
  {\bibfnamefont{F.~H.}\ \bibnamefont{Baumann}}, \bibinfo {author}
  {\bibfnamefont{K.}~\bibnamefont{Evans-Lutterodt}},\ and\ \bibinfo {author}
  {\bibfnamefont{G.}~\bibnamefont{Timp}},\ }%
  \bibfield{journal}{%
  \bibinfo {journal} {Nature}\ }%
  \textbf{\bibinfo {volume} {399}},\ \bibinfo {pages} {758} (\bibinfo {year}
  {1999})%
  \bibAnnoteFile{NoStop}{muller1999a}%
\bibitem{meindl2001a}%
  \BibitemOpen
  \bibfield{author}{%
  \bibinfo {author} {\bibfnamefont{J.~D.}\ \bibnamefont{Meindl}}, \bibinfo
  {author} {\bibfnamefont{Q.}~\bibnamefont{Chen}},\ and\ \bibinfo {author}
  {\bibfnamefont{J.~A.}\ \bibnamefont{Davis}},\ }%
  \bibfield{journal}{%
  \bibinfo {journal} {Science}\ }%
  \textbf{\bibinfo {volume} {293}},\ \bibinfo {pages} {2044} (\bibinfo {year}
  {2001})%
  \bibAnnoteFile{NoStop}{meindl2001a}%
\bibitem{aviram1974a}%
  \BibitemOpen
  \bibfield{author}{%
  \bibinfo {author} {\bibfnamefont{A.}~\bibnamefont{Aviram}}\ and\ \bibinfo
  {author} {\bibfnamefont{M.~A.}\ \bibnamefont{Ratner}},\ }%
  \bibfield{journal}{%
  \bibinfo {journal} {Chem. Phys. Lett.}\ }%
  \textbf{\bibinfo {volume} {29}},\ \bibinfo {pages} {277} (\bibinfo {year}
  {1974})%
  \bibAnnoteFile{NoStop}{aviram1974a}%
\bibitem{joachim2000a}%
  \BibitemOpen
  \bibfield{author}{%
  \bibinfo {author} {\bibfnamefont{C.}~\bibnamefont{Joachim}}, \bibinfo
  {author} {\bibfnamefont{J.~K.}\ \bibnamefont{Gimzewski}},\ and\ \bibinfo
  {author} {\bibfnamefont{A.}~\bibnamefont{Aviram}},\ }%
  \bibfield{journal}{%
  \bibinfo {journal} {Nature}\ }%
  \textbf{\bibinfo {volume} {408}},\ \bibinfo {pages} {541} (\bibinfo {year}
  {2000})%
  \bibAnnoteFile{NoStop}{joachim2000a}%
\bibitem{kahn1988a}%
  \BibitemOpen
  \bibfield{author}{%
  \bibinfo {author} {\bibfnamefont{O.}~\bibnamefont{Kahn}}\ and\ \bibinfo
  {author} {\bibfnamefont{J.~P.}\ \bibnamefont{Launay}},\ }%
  \bibfield{journal}{%
  \bibinfo {journal} {Chemtronics}\ }%
  \textbf{\bibinfo {volume} {3}},\ \bibinfo {pages} {140} (\bibinfo {year}
  {1988})%
  \bibAnnoteFile{NoStop}{kahn1988a}%
\bibitem{raymo2002a}%
  \BibitemOpen
  \bibfield{author}{%
  \bibinfo {author} {\bibfnamefont{F.~M.}\ \bibnamefont{Raymo}},\ }%
  \bibfield{journal}{%
  \bibinfo {journal} {Adv. Mater.}\ }%
  \textbf{\bibinfo {volume} {14}},\ \bibinfo {pages} {401} (\bibinfo {year}
  {2002})%
  \bibAnnoteFile{NoStop}{raymo2002a}%
\bibitem{liu2010a}%
  \BibitemOpen
  \bibfield{author}{%
  \bibinfo {author} {\bibfnamefont{Y.}~\bibnamefont{Liu}}, \bibinfo {author}
  {\bibfnamefont{A.}~\bibnamefont{Offenh{\"a}usser}},\ and\ \bibinfo {author}
  {\bibfnamefont{D.}~\bibnamefont{Mayer}},\ }%
  \bibfield{journal}{%
  \bibinfo {journal} {Angew. Chem. Int. Ed.}\ }%
  \textbf{\bibinfo {volume} {49}},\ \bibinfo {pages} {2595} (\bibinfo {year}
  {2010})%
  \bibAnnoteFile{NoStop}{liu2010a}%
\bibitem{puntoriero2011a}%
  \BibitemOpen
  \bibfield{author}{%
  \bibinfo {author} {\bibfnamefont{F.}~\bibnamefont{Puntoriero}}, \bibinfo
  {author} {\bibfnamefont{F.}~\bibnamefont{Nastasi}}, \bibinfo {author}
  {\bibfnamefont{T.}~\bibnamefont{Bura}}, \bibinfo {author}
  {\bibfnamefont{R.}~\bibnamefont{Ziessel}}, \bibinfo {author}
  {\bibfnamefont{S.}~\bibnamefont{Campagna}},\ and\ \bibinfo {author}
  {\bibfnamefont{A.}~\bibnamefont{Giannetto}},\ }%
  \bibfield{journal}{%
  \bibinfo {journal} {New J. Chem.}\ }%
  \textbf{\bibinfo {volume} {35}},\ \bibinfo {pages} {948} (\bibinfo {year}
  {2011})%
  \bibAnnoteFile{NoStop}{puntoriero2011a}%
\bibitem{renaud2011b}%
  \BibitemOpen
  \bibfield{author}{%
  \bibinfo {author} {\bibfnamefont{N.}~\bibnamefont{Renaud}}, \bibinfo {author}
  {\bibfnamefont{M.}~\bibnamefont{Hliwa}},\ and\ \bibinfo {author}
  {\bibfnamefont{C.}~\bibnamefont{Joachim}},\ }%
  \bibfield{journal}{%
  \bibinfo {journal} {Phys. Chem. Chem. Phys.}\ }%
  \textbf{\bibinfo {volume} {13}},\ \bibinfo {pages} {14404} (\bibinfo {year}
  {2011})%
  \bibAnnoteFile{NoStop}{renaud2011b}%
\bibitem{kawai2012a}%
  \BibitemOpen
  \bibfield{author}{%
  \bibinfo {author} {\bibfnamefont{H.}~\bibnamefont{Kawai}}, \bibinfo {author}
  {\bibfnamefont{F.}~\bibnamefont{Ample}}, \bibinfo {author}
  {\bibfnamefont{Q.}~\bibnamefont{Wang}}, \bibinfo {author}
  {\bibfnamefont{Y.~K.}\ \bibnamefont{Yeo}}, \bibinfo {author}
  {\bibfnamefont{M.}~\bibnamefont{Saeys}},\ and\ \bibinfo {author}
  {\bibfnamefont{C.}~\bibnamefont{Joachim}},\ }%
  \bibfield{journal}{%
  \bibinfo {journal} {J. Phys.: Condens. Matter}\ }%
  \textbf{\bibinfo {volume} {24}},\ \bibinfo {pages} {095011} (\bibinfo {year}
  {2012})%
  \bibAnnoteFile{NoStop}{kawai2012a}%
\bibitem{shen1995a}%
  \BibitemOpen
  \bibfield{author}{%
  \bibinfo {author} {\bibfnamefont{T.-C.}\ \bibnamefont{Shen}}, \bibinfo
  {author} {\bibfnamefont{C.}~\bibnamefont{Wang}}, \bibinfo {author}
  {\bibfnamefont{G.~C.}\ \bibnamefont{Abeln}}, \bibinfo {author}
  {\bibfnamefont{J.~R.}\ \bibnamefont{Tucker}}, \bibinfo {author}
  {\bibfnamefont{J.~W.}\ \bibnamefont{Lyding}}, \bibinfo {author}
  {\bibfnamefont{P.}~\bibnamefont{Avouris}},\ and\ \bibinfo {author}
  {\bibfnamefont{R.~E.}\ \bibnamefont{Walkup}},\ }%
  \bibfield{journal}{%
  \bibinfo {journal} {Science}\ }%
  \textbf{\bibinfo {volume} {268}},\ \bibinfo {pages} {1590} (\bibinfo {year}
  {1995})%
  \bibAnnoteFile{NoStop}{shen1995a}%
\bibitem{hosaka1995a}%
  \BibitemOpen
  \bibfield{author}{%
  \bibinfo {author} {\bibfnamefont{S.}~\bibnamefont{Hosaka}}, \bibinfo {author}
  {\bibfnamefont{S.}~\bibnamefont{Hosoki}}, \bibinfo {author}
  {\bibfnamefont{T.}~\bibnamefont{Hasegawa}}, \bibinfo {author}
  {\bibfnamefont{H.}~\bibnamefont{Koyanagi}}, \bibinfo {author}
  {\bibfnamefont{T.}~\bibnamefont{Shintani}},\ and\ \bibinfo {author}
  {\bibfnamefont{M.}~\bibnamefont{Miyamoto}},\ }%
  \bibfield{journal}{%
  \bibinfo {journal} {J. Vac. Sci. Technol. B}\ }%
  \textbf{\bibinfo {volume} {13}},\ \bibinfo {pages} {2813} (\bibinfo {year}
  {1995})%
  \bibAnnoteFile{NoStop}{hosaka1995a}%
\bibitem{hitosugi1999a}%
  \BibitemOpen
  \bibfield{author}{%
  \bibinfo {author} {\bibfnamefont{T.}~\bibnamefont{Hitosugi}}, \bibinfo
  {author} {\bibfnamefont{S.}~\bibnamefont{Heike}}, \bibinfo {author}
  {\bibfnamefont{T.}~\bibnamefont{Onogi}}, \bibinfo {author}
  {\bibfnamefont{T.}~\bibnamefont{Hashizume}}, \bibinfo {author}
  {\bibfnamefont{S.}~\bibnamefont{Watanabe}}, \bibinfo {author}
  {\bibfnamefont{Z.-Q.}\ \bibnamefont{Li}}, \bibinfo {author}
  {\bibfnamefont{K.}~\bibnamefont{Ohno}}, \bibinfo {author}
  {\bibfnamefont{Y.}~\bibnamefont{Kawazoe}}, \bibinfo {author}
  {\bibfnamefont{T.}~\bibnamefont{Hasegawa}},\ and\ \bibinfo {author}
  {\bibfnamefont{K.}~\bibnamefont{Kitazawa}},\ }%
  \bibfield{journal}{%
  \bibinfo {journal} {Phys. Rev. Lett.}\ }%
  \textbf{\bibinfo {volume} {82}},\ \bibinfo {pages} {4034} (\bibinfo {year}
  {1999})%
  \bibAnnoteFile{NoStop}{hitosugi1999a}%
\bibitem{soukiassian2003a}%
  \BibitemOpen
  \bibfield{author}{%
  \bibinfo {author} {\bibfnamefont{L.}~\bibnamefont{Soukiassian}}, \bibinfo
  {author} {\bibfnamefont{A.~J.}\ \bibnamefont{Mayne}}, \bibinfo {author}
  {\bibfnamefont{M.}~\bibnamefont{Carbone}},\ and\ \bibinfo {author}
  {\bibfnamefont{G.}~\bibnamefont{Dujardin}},\ }%
  \bibfield{journal}{%
  \bibinfo {journal} {Surf. Sci.}\ }%
  \textbf{\bibinfo {volume} {528}},\ \bibinfo {pages} {121} (\bibinfo {year}
  {2003})%
  \bibAnnoteFile{NoStop}{soukiassian2003a}%
\bibitem{hallam2007a}%
  \BibitemOpen
  \bibfield{author}{%
  \bibinfo {author} {\bibfnamefont{T.}~\bibnamefont{Hallam}}, \bibinfo {author}
  {\bibfnamefont{T.~C.~G.}\ \bibnamefont{Reusch}}, \bibinfo {author}
  {\bibfnamefont{L.}~\bibnamefont{Oberbeck}}, \bibinfo {author}
  {\bibfnamefont{N.~J.}\ \bibnamefont{Curson}},\ and\ \bibinfo {author}
  {\bibfnamefont{M.~Y.}\ \bibnamefont{Simmons}},\ }%
  \bibfield{journal}{%
  \bibinfo {journal} {J. Appl. Phys.}\ }%
  \textbf{\bibinfo {volume} {101}},\ \bibinfo {pages} {034305} (\bibinfo {year}
  {2007})%
  \bibAnnoteFile{NoStop}{hallam2007a}%
\bibitem{haider2009a}%
  \BibitemOpen
  \bibfield{author}{%
  \bibinfo {author} {\bibfnamefont{M.~B.}\ \bibnamefont{Haider}}, \bibinfo
  {author} {\bibfnamefont{J.~L.}\ \bibnamefont{Pitters}}, \bibinfo {author}
  {\bibfnamefont{G.~A.}\ \bibnamefont{DiLabio}}, \bibinfo {author}
  {\bibfnamefont{L.}~\bibnamefont{Livadaru}}, \bibinfo {author}
  {\bibfnamefont{J.~Y.}\ \bibnamefont{Mutus}},\ and\ \bibinfo {author}
  {\bibfnamefont{R.~A.}\ \bibnamefont{Wolkow}},\ }%
  \bibfield{journal}{%
  \bibinfo {journal} {Phys. Rev. Lett.}\ }%
  \textbf{\bibinfo {volume} {102}},\ \bibinfo {pages} {046805} (\bibinfo {year}
  {2009})%
  \bibAnnoteFile{NoStop}{haider2009a}%
\bibitem{pitters2011a}%
  \BibitemOpen
  \bibfield{author}{%
  \bibinfo {author} {\bibfnamefont{J.~L.}\ \bibnamefont{Pitters}}, \bibinfo
  {author} {\bibfnamefont{L.}~\bibnamefont{Livadaru}}, \bibinfo {author}
  {\bibfnamefont{M.~B.}\ \bibnamefont{Haider}},\ and\ \bibinfo {author}
  {\bibfnamefont{R.~A.}\ \bibnamefont{Wolkow}},\ }%
  \bibfield{journal}{%
  \bibinfo {journal} {J. Chem. Phys.}\ }%
  \textbf{\bibinfo {volume} {134}},\ \bibinfo {pages} {064712} (\bibinfo {year}
  {2011})%
  \bibAnnoteFile{NoStop}{pitters2011a}%
\bibitem{doumergue1999a}%
  \BibitemOpen
  \bibfield{author}{%
  \bibinfo {author} {\bibfnamefont{P.}~\bibnamefont{Doumergue}}, \bibinfo
  {author} {\bibfnamefont{L.}~\bibnamefont{Pizzagalli}}, \bibinfo {author}
  {\bibfnamefont{C.}~\bibnamefont{Joachim}}, \bibinfo {author}
  {\bibfnamefont{A.}~\bibnamefont{Altibelli}},\ and\ \bibinfo {author}
  {\bibfnamefont{A.}~\bibnamefont{Baratoff}},\ }%
  \bibfield{journal}{%
  \bibinfo {journal} {Phys. Rev. B}\ }%
  \textbf{\bibinfo {volume} {59}},\ \bibinfo {pages} {15910} (\bibinfo {year}
  {1999})%
  \bibAnnoteFile{NoStop}{doumergue1999a}%
\bibitem{kawai2010a}%
  \BibitemOpen
  \bibfield{author}{%
  \bibinfo {author} {\bibfnamefont{H.}~\bibnamefont{Kawai}}, \bibinfo {author}
  {\bibfnamefont{Y.~K.}\ \bibnamefont{Yeo}}, \bibinfo {author}
  {\bibfnamefont{M.}~\bibnamefont{Saeys}},\ and\ \bibinfo {author}
  {\bibfnamefont{C.}~\bibnamefont{Joachim}},\ }%
  \bibfield{journal}{%
  \bibinfo {journal} {Phys. Rev. B}\ }%
  \textbf{\bibinfo {volume} {81}},\ \bibinfo {pages} {195316} (\bibinfo {year}
  {2010})%
  \bibAnnoteFile{NoStop}{kawai2010a}%
\bibitem{watanabe1996a}%
  \BibitemOpen
  \bibfield{author}{%
  \bibinfo {author} {\bibfnamefont{S.}~\bibnamefont{Watanabe}}, \bibinfo
  {author} {\bibfnamefont{Y.~A.}\ \bibnamefont{Ono}}, \bibinfo {author}
  {\bibfnamefont{T.}~\bibnamefont{Hashizume}},\ and\ \bibinfo {author}
  {\bibfnamefont{Y.}~\bibnamefont{Wada}},\ }%
  \bibfield{journal}{%
  \bibinfo {journal} {Phys. Rev. B}\ }%
  \textbf{\bibinfo {volume} {54}},\ \bibinfo {pages} {R17308} (\bibinfo {year}
  {1996})%
  \bibAnnoteFile{NoStop}{watanabe1996a}%
\bibitem{watanabe1997a}%
  \BibitemOpen
  \bibfield{author}{%
  \bibinfo {author} {\bibfnamefont{S.}~\bibnamefont{Watanabe}}, \bibinfo
  {author} {\bibfnamefont{Y.~A.}\ \bibnamefont{Ono}}, \bibinfo {author}
  {\bibfnamefont{T.}~\bibnamefont{Hashizume}},\ and\ \bibinfo {author}
  {\bibfnamefont{Y.}~\bibnamefont{Wada}},\ }%
  \bibfield{journal}{%
  \bibinfo {journal} {Surf. Sci.}\ }%
  \textbf{\bibinfo {volume} {386}},\ \bibinfo {pages} {340} (\bibinfo {year}
  {1997})%
  \bibAnnoteFile{NoStop}{watanabe1997a}%
\bibitem{cho2002a}%
  \BibitemOpen
  \bibfield{author}{%
  \bibinfo {author} {\bibfnamefont{J.-H.}\ \bibnamefont{Cho}}\ and\ \bibinfo
  {author} {\bibfnamefont{L.}~\bibnamefont{Kleinman}},\ }%
  \bibfield{journal}{%
  \bibinfo {journal} {Phys. Rev. B}\ }%
  \textbf{\bibinfo {volume} {66}},\ \bibinfo {pages} {235405} (\bibinfo {year}
  {2002})%
  \bibAnnoteFile{NoStop}{cho2002a}%
\bibitem{bird2003}%
  \BibitemOpen
  \bibfield{author}{%
  \bibinfo {author} {\bibfnamefont{C.~F.}\ \bibnamefont{Bird}}\ and\ \bibinfo
  {author} {\bibfnamefont{D.~R.}\ \bibnamefont{Bowler}},\ }%
  \bibfield{journal}{%
  \bibinfo {journal} {Surf. Sci.}\ }%
  \textbf{\bibinfo {volume} {531}},\ \bibinfo {pages} {L351} (\bibinfo {year}
  {2003})%
  \bibAnnoteFile{NoStop}{bird2003}%
\bibitem{cakmak2003a}%
  \BibitemOpen
  \bibfield{author}{%
  \bibinfo {author} {\bibfnamefont{M.}~\bibnamefont{{\c C}akmak}}\ and\
  \bibinfo {author} {\bibfnamefont{G.~P.}\ \bibnamefont{Srivastava}},\ }%
  \bibfield{journal}{%
  \bibinfo {journal} {Surf. Sci.}\ }%
  \textbf{\bibinfo {volume} {532-535}},\ \bibinfo {pages} {556} (\bibinfo
  {year} {2003})%
  \bibAnnoteFile{NoStop}{cakmak2003a}%
\bibitem{lee2008a}%
  \BibitemOpen
  \bibfield{author}{%
  \bibinfo {author} {\bibfnamefont{J.~Y.}\ \bibnamefont{Lee}}, \bibinfo
  {author} {\bibfnamefont{J.-H.}\ \bibnamefont{Choi}},\ and\ \bibinfo {author}
  {\bibfnamefont{J.-H.}\ \bibnamefont{Cho}},\ }%
  \bibfield{journal}{%
  \bibinfo {journal} {Phys. Rev. B}\ }%
  \textbf{\bibinfo {volume} {78}},\ \bibinfo {pages} {081303} (\bibinfo {year}
  {2008})%
  \bibAnnoteFile{NoStop}{lee2008a}%
\bibitem{lee2009a}%
  \BibitemOpen
  \bibfield{author}{%
  \bibinfo {author} {\bibfnamefont{J.~Y.}\ \bibnamefont{Lee}}, \bibinfo
  {author} {\bibfnamefont{J.-H.}\ \bibnamefont{Cho}},\ and\ \bibinfo {author}
  {\bibfnamefont{Z.}~\bibnamefont{Zhang}},\ }%
  \bibfield{journal}{%
  \bibinfo {journal} {Phys. Rev. B}\ }%
  \textbf{\bibinfo {volume} {80}},\ \bibinfo {pages} {155329} (\bibinfo {year}
  {2009})%
  \bibAnnoteFile{NoStop}{lee2009a}%
\bibitem{lee2011a}%
  \BibitemOpen
  \bibfield{author}{%
  \bibinfo {author} {\bibfnamefont{J.-H.}\ \bibnamefont{Lee}}\ and\ \bibinfo
  {author} {\bibfnamefont{J.-H.}\ \bibnamefont{Cho}},\ }%
  \bibfield{journal}{%
  \bibinfo {journal} {Surf. Sci.}\ }%
  \textbf{\bibinfo {volume} {605}},\ \bibinfo {pages} {L13} (\bibinfo {year}
  {2011})%
  \bibAnnoteFile{NoStop}{lee2011a}%
\bibitem{nous2011a}%
  \BibitemOpen
  \bibfield{author}{%
  \bibinfo {author} {\bibfnamefont{R.}~\bibnamefont{Robles}}, \bibinfo {author}
  {\bibfnamefont{M.}~\bibnamefont{Kepenekian}}, \bibinfo {author}
  {\bibfnamefont{S.}~\bibnamefont{Monturet}}, \bibinfo {author}
  {\bibfnamefont{C.}~\bibnamefont{Joachim}},\ and\ \bibinfo {author}
  {\bibfnamefont{N.}~\bibnamefont{Lorente}},\ }%
  \bibinfo {journal} {ArXiv:1203.2548v1}%
  \bibAnnoteFile{NoStop}{nous2011a}%
\bibitem{marzari1997a}%
  \BibitemOpen
\bibfield{journal}{%
    }%
  \bibfield{author}{%
  \bibinfo {author} {\bibfnamefont{N.}~\bibnamefont{Marzari}}\ and\ \bibinfo
  {author} {\bibfnamefont{D.}~\bibnamefont{Vanderbilt}},\ }%
  \bibfield{journal}{%
  \bibinfo {journal} {Phys. Rev. B}\ }%
  \textbf{\bibinfo {volume} {56}},\ \bibinfo {pages} {12847} (\bibinfo {year}
  {1997})%
  \bibAnnoteFile{NoStop}{marzari1997a}%
\bibitem{soler2002a}%
  \BibitemOpen
  \bibfield{author}{%
  \bibinfo {author} {\bibfnamefont{J.~M.}\ \bibnamefont{Soler}}, \bibinfo
  {author} {\bibfnamefont{E.}~\bibnamefont{Artacho}}, \bibinfo {author}
  {\bibfnamefont{J.~D.}\ \bibnamefont{Gale}}, \bibinfo {author}
  {\bibfnamefont{A.}~\bibnamefont{Garc{\'i}a}}, \bibinfo {author}
  {\bibfnamefont{J.}~\bibnamefont{Junquera}}, \bibinfo {author}
  {\bibfnamefont{P.}~\bibnamefont{Ordej{\'o}n}},\ and\ \bibinfo {author}
  {\bibfnamefont{D.}~\bibnamefont{S{\'a}nchez-Portal}},\ }%
  \bibfield{journal}{%
  \bibinfo {journal} {J. Phys.: Condens. Matter}\ }%
  \textbf{\bibinfo {volume} {14}},\ \bibinfo {pages} {2745} (\bibinfo {year}
  {2002})%
  \bibAnnoteFile{NoStop}{soler2002a}%
\bibitem{artacho2008a}%
  \BibitemOpen
  \bibfield{author}{%
  \bibinfo {author} {\bibfnamefont{E.}~\bibnamefont{Artacho}}, \bibinfo
  {author} {\bibfnamefont{E.}~\bibnamefont{Anglada}}, \bibinfo {author}
  {\bibfnamefont{O.}~\bibnamefont{Di{\'e}guez}}, \bibinfo {author}
  {\bibfnamefont{J.~D.}\ \bibnamefont{Gale}}, \bibinfo {author}
  {\bibfnamefont{A.}~\bibnamefont{Garc{\'i}a}}, \bibinfo {author}
  {\bibfnamefont{J.}~\bibnamefont{Junquera}}, \bibinfo {author}
  {\bibfnamefont{R.~M.}\ \bibnamefont{Martin}}, \bibinfo {author}
  {\bibfnamefont{P.}~\bibnamefont{Ordej{\'o}n}}, \bibinfo {author}
  {\bibfnamefont{J.~M.}\ \bibnamefont{Pruneda}}, \bibinfo {author}
  {\bibfnamefont{D.}~\bibnamefont{S{\'a}nchez-Portal}},\ and\ \bibinfo {author}
  {\bibfnamefont{J.~M.}\ \bibnamefont{Soler}},\ }%
  \bibfield{journal}{%
  \bibinfo {journal} {J. Phys.: Condens. Matter}\ }%
  \textbf{\bibinfo {volume} {20}},\ \bibinfo {pages} {064208} (\bibinfo {year}
  {2008})%
  \bibAnnoteFile{NoStop}{artacho2008a}%
\bibitem{perdew1996a}%
  \BibitemOpen
  \bibfield{author}{%
  \bibinfo {author} {\bibfnamefont{J.~P.}\ \bibnamefont{Perdew}}, \bibinfo
  {author} {\bibfnamefont{K.}~\bibnamefont{Burke}},\ and\ \bibinfo {author}
  {\bibfnamefont{M.}~\bibnamefont{Ernzerhof}},\ }%
  \bibfield{journal}{%
  \bibinfo {journal} {Phys. Rev. Lett.}\ }%
  \textbf{\bibinfo {volume} {77}},\ \bibinfo {pages} {3865} (\bibinfo {year}
  {1996})%
  \bibAnnoteFile{NoStop}{perdew1996a}%
\bibitem{troullier1991a}%
  \BibitemOpen
  \bibfield{author}{%
  \bibinfo {author} {\bibfnamefont{N.}~\bibnamefont{Troullier}}\ and\ \bibinfo
  {author} {\bibfnamefont{J.~L.}\ \bibnamefont{Martins}},\ }%
  \bibfield{journal}{%
  \bibinfo {journal} {Phys. Rev. B}\ }%
  \textbf{\bibinfo {volume} {43}},\ \bibinfo {pages} {1993} (\bibinfo {year}
  {1991})%
  \bibAnnoteFile{NoStop}{troullier1991a}%
\bibitem{artacho1999a}%
  \BibitemOpen
  \bibfield{author}{%
  \bibinfo {author} {\bibfnamefont{E.}~\bibnamefont{Artacho}}, \bibinfo
  {author} {\bibfnamefont{D.}~\bibnamefont{S{\'a}nchez-Portal}}, \bibinfo
  {author} {\bibfnamefont{P.}~\bibnamefont{Ordej{\'o}n}}, \bibinfo {author}
  {\bibfnamefont{A.}~\bibnamefont{Garc{\'i}a}},\ and\ \bibinfo {author}
  {\bibfnamefont{J.~M.}\ \bibnamefont{Soler}},\ }%
  \bibfield{journal}{%
  \bibinfo {journal} {phys. stat. sol. (b)}\ }%
  \textbf{\bibinfo {volume} {215}},\ \bibinfo {pages} {809} (\bibinfo {year}
  {1999})%
  \bibAnnoteFile{NoStop}{artacho1999a}%
\bibitem{brandbyge2002a}%
  \BibitemOpen
  \bibfield{author}{%
  \bibinfo {author} {\bibfnamefont{M.}~\bibnamefont{Brandbyge}}, \bibinfo
  {author} {\bibfnamefont{J.-L.}\ \bibnamefont{Mozos}}, \bibinfo {author}
  {\bibfnamefont{P.}~\bibnamefont{Ordej{\'o}n}}, \bibinfo {author}
  {\bibfnamefont{J.}~\bibnamefont{Taylor}},\ and\ \bibinfo {author}
  {\bibfnamefont{K.}~\bibnamefont{Stokbro}},\ }%
  \bibfield{journal}{%
  \bibinfo {journal} {Phys. Rev. B}\ }%
  \textbf{\bibinfo {volume} {65}},\ \bibinfo {pages} {165401} (\bibinfo {year}
  {2002})%
  \bibAnnoteFile{NoStop}{brandbyge2002a}%
\bibitem{mostofi2008a}%
  \BibitemOpen
  \bibfield{author}{%
  \bibinfo {author} {\bibfnamefont{A.~A.}\ \bibnamefont{Mostofi}}, \bibinfo
  {author} {\bibfnamefont{J.~R.}\ \bibnamefont{Yates}}, \bibinfo {author}
  {\bibfnamefont{Y.-S.}\ \bibnamefont{Lee}}, \bibinfo {author}
  {\bibfnamefont{I.}~\bibnamefont{Souza}}, \bibinfo {author}
  {\bibfnamefont{D.}~\bibnamefont{Vanderbilt}},\ and\ \bibinfo {author}
  {\bibfnamefont{N.}~\bibnamefont{Marzari}},\ }%
  \bibfield{journal}{%
  \bibinfo {journal} {Comput. Phys. Commun.}\ }%
  \textbf{\bibinfo {volume} {178}},\ \bibinfo {pages} {685} (\bibinfo {year}
  {2008})%
  \bibAnnoteFile{NoStop}{mostofi2008a}%
\bibitem{korytar2010a}%
  \BibitemOpen
  \bibfield{author}{%
  \bibinfo {author} {\bibfnamefont{R.}~\bibnamefont{Koryt{\'a}r}}, \bibinfo
  {author} {\bibfnamefont{J.~M.}\ \bibnamefont{Pruneda}}, \bibinfo {author}
  {\bibfnamefont{J.}~\bibnamefont{Junquera}}, \bibinfo {author}
  {\bibfnamefont{P.}~\bibnamefont{Ordej{\'o}n}},\ and\ \bibinfo {author}
  {\bibfnamefont{N.}~\bibnamefont{Lorente}},\ }%
  \bibfield{journal}{%
  \bibinfo {journal} {J. Phys.: Condens. Matter}\ }%
  \textbf{\bibinfo {volume} {22}},\ \bibinfo {pages} {385601} (\bibinfo {year}
  {2010})%
  \bibAnnoteFile{NoStop}{korytar2010a}%
\bibitem{korytar2011a}%
  \BibitemOpen
  \bibfield{author}{%
  \bibinfo {author} {\bibfnamefont{R.}~\bibnamefont{Koryt{\'a}r}}\ and\
  \bibinfo {author} {\bibfnamefont{N.}~\bibnamefont{Lorente}},\ }%
  \bibfield{journal}{%
  \bibinfo {journal} {J. Phys.: Condens. Matter}\ }%
  \textbf{\bibinfo {volume} {23}},\ \bibinfo {pages} {355009} (\bibinfo {year}
  {2011})%
  \bibAnnoteFile{NoStop}{korytar2011a}%
\bibitem{paulsson2007a}%
  \BibitemOpen
  \bibfield{author}{%
  \bibinfo {author} {\bibfnamefont{M.}~\bibnamefont{Paulsson}}\ and\ \bibinfo
  {author} {\bibfnamefont{M.}~\bibnamefont{Brandbyge}},\ }%
  \bibfield{journal}{%
  \bibinfo {journal} {Physical Review B}\ }%
  \textbf{\bibinfo {volume} {76}},\ \bibinfo {pages} {115117} (\bibinfo {year}
  {2007})%
  \bibAnnoteFile{NoStop}{paulsson2007a}%
\bibitem{frederiksen2007a}%
  \BibitemOpen
  \bibfield{author}{%
  \bibinfo {author} {\bibfnamefont{T.}~\bibnamefont{Frederiksen}}, \bibinfo
  {author} {\bibfnamefont{M.}~\bibnamefont{Paulsson}}, \bibinfo {author}
  {\bibfnamefont{M.}~\bibnamefont{Brandbyge}},\ and\ \bibinfo {author}
  {\bibfnamefont{A.P.}~\bibnamefont{Jauho}},\ }%
  \bibfield{journal}{%
  \bibinfo {journal} {Physical Review B}\ }%
  \textbf{\bibinfo {volume} {75}},\ \bibinfo {pages} {205413} (\bibinfo {year}
  {2007})%
  \bibAnnoteFile{NoStop}{frederiksen2007a}%
\bibitem{liang1998a}%
  \BibitemOpen
  \bibfield{author}{%
  \bibinfo {author} {\bibfnamefont{G.}~\bibnamefont{Liang}}, \bibinfo {author}
  {\bibfnamefont{Y.~A.}\ \bibnamefont{Lin}}, \bibinfo {author}
  {\bibfnamefont{D.~Z.}\ \bibnamefont{Ting}},\ and\ \bibinfo {author}
  {\bibfnamefont{Y.}~\bibnamefont{Chang}},\ }%
  \bibfield{journal}{%
  \bibinfo {journal} {{VLSI} Design}\ }%
  \textbf{\bibinfo {volume} {8}},\ \bibinfo {pages} {507} (\bibinfo {year}
  {1998}),\ ISSN \bibinfo {issn} {{1065-514X}, 1563-5171}%
  \bibAnnoteFile{NoStop}{liang1998a}%
\bibitem{datta_book}%
  \BibitemOpen
  \bibfield{author}{%
  \bibinfo {author} {\bibfnamefont{S.}~\bibnamefont{Datta}},\ }%
  \emph{\bibinfo {title} {Electronic Transport in Mesoscopic Systems}}\
  (\bibinfo {publisher} {Cambridge University Press},\ \bibinfo {year} {2007})%
  \bibAnnoteFile{NoStop}{datta_book}%
\bibitem{bowler2001a}%
  \BibitemOpen
  \bibfield{author}{%
  \bibinfo {author} {\bibfnamefont{D.~R.}\ \bibnamefont{Bowler}}\ and\ \bibinfo
  {author} {\bibfnamefont{A.~J.}\ \bibnamefont{Fisher}},\ }%
  \bibfield{journal}{%
  \bibinfo {journal} {Phys. Rev. B}\ }%
  \textbf{\bibinfo {volume} {63}},\ \bibinfo {pages} {035310} (\bibinfo {year}
  {2000})%
  \bibAnnoteFile{NoStop}{bowler2001a}%
\bibitem{todorovic2002a}%
  \BibitemOpen
  \bibfield{author}{%
  \bibinfo {author} {\bibfnamefont{M.}~\bibnamefont{Todorovic}}, \bibinfo
  {author} {\bibfnamefont{A.~J.}\ \bibnamefont{Fisher}},\ and\ \bibinfo
  {author} {\bibfnamefont{D.~R.}\ \bibnamefont{Bowler}},\ }%
  \bibfield{journal}{%
  \bibinfo {journal} {J. Phys.: Condens. Matter}\ }%
  \textbf{\bibinfo {volume} {14}},\ \bibinfo {pages} {L749–L755} (\bibinfo
  {year} {2002})%
  \bibAnnoteFile{NoStop}{todorovic2002a}%
\bibitem{futuro}%
  \BibitemOpen
  \bibfield{author}{%
  \bibinfo {author} {\bibfnamefont{S.}~\bibnamefont{Monturet}}, \bibinfo
  {author} {\bibfnamefont{M.}~\bibnamefont{Kepenekian}}, \bibinfo {author}
  {\bibfnamefont{R.}~\bibnamefont{Robles}}, \bibinfo {author}
  {\bibfnamefont{N.}~\bibnamefont{Lorente}},\ and\ \bibinfo {author}
  {\bibfnamefont{C.}~\bibnamefont{Joachim}},\ }%
  \bibinfo {journal} {unpublished}%
  \bibAnnoteFile{NoStop}{futuro}%
\end{thebibliography}%
\bibliographystyle{apsrev4-1}

\end{document}